\documentclass[aps,prl,preprint,nopacs,superscriptaddress]{revtex4}

\usepackage{graphicx}
\usepackage{verbatim}
\usepackage{mathrsfs}
\usepackage{array}
\newcolumntype{P}[1]{>{\centering\arraybackslash}p{#1}}
\usepackage{color}

\usepackage{ulem}

\usepackage{amsmath,amsfonts,amssymb}

\def\3{2.8in}    
\def\2{2.5in}
\def\4{3.0in}

\def \beq {\begin{equation}}
\def \eeq {\end{equation}}

\begin{document}

\title{Weyl, Dirac and high-fold chiral fermions \\ in topological quantum materials}

\author{M. Zahid Hasan$^{\dag}$\footnote[0]{$^{\dag}$Corresponding author (email):mzhasan@princeton.edu }}
\affiliation {Laboratory for Topological Quantum Matter and Advanced Spectroscopy (B7), Department of Physics, Princeton University, Princeton, New Jersey 08544, USA}
\affiliation{Materials Sciences Division, Lawrence Berkeley National Laboratory, Berkeley, CA 94720, USA}
\affiliation{Princeton Institute for Science and Technology of Materials, Princeton University, Princeton, NJ, 08544, USA}

\author{Guoqing Chang $^{\dag}$\footnote[0]{$^{\dag}$Corresponding author (email):guoqing.chang@ntu.edu.sg }}
\affiliation {Laboratory for Topological Quantum Matter and Advanced Spectroscopy (B7), Department of Physics, Princeton University, Princeton, New Jersey 08544, USA}
\affiliation {Division of Physics and Applied Physics, School of Physical and Mathematical Sciences, Nanyang Technological University, Singapore 637371, Singapore}

\author{Ilya Belopolski}
\affiliation {Laboratory for Topological Quantum Matter and Advanced Spectroscopy (B7), Department of Physics, Princeton University, Princeton, New Jersey 08544, USA}

\author{Guang Bian}
\affiliation {Laboratory for Topological Quantum Matter and Advanced Spectroscopy (B7), Department of Physics, Princeton University, Princeton, New Jersey 08544, USA}
\affiliation {Department of Physics and Astronomy, University of Missouri, Columbia, Missouri 65211, USA}

\author{Su-Yang Xu}
\affiliation {Laboratory for Topological Quantum Matter and Advanced Spectroscopy (B7), Department of Physics, Princeton University, Princeton, New Jersey 08544, USA}
\affiliation {Department of Chemistry and Chemical Biology, Harvard University, Cambridge, MA 02138, USA}

\author{Jia-Xin Yin}\affiliation {Laboratory for Topological Quantum Matter and Advanced Spectroscopy (B7), Department of Physics, Princeton University, Princeton, New Jersey 08544, USA}
\begin{abstract}

Quantum materials hosting Weyl fermions have opened a new era of research in condensed matter physics. First proposed in 1929 in the context of particle physics, Weyl fermions have yet to be observed as elementary particles. In 2015, Weyl fermions were detected as collective electronic excitations in the strong spin-orbit coupled material tantalum arsenide, TaAs. This discovery was followed by a flurry of experimental and theoretical explorations of Weyl phenomena in materials. Weyl materials naturally lend themselves to the exploration of the topological index associated with Weyl fermions and their divergent Berry curvature field, as well as the topological bulk-boundary correspondence giving rise to protected conducting surface states. Here, we review the broader class of Weyl topological phenomena in materials, starting with the observation of emergent Weyl fermions in the bulk and of Fermi arc states on the surface of the TaAs family of crystals by photoemission spectroscopy. We then discuss some of the exotic optical and magnetic responses observed in these materials, as well as the progress in developing some of the related chiral materials. We discuss the conceptual development of high-fold chiral fermions, which generalize Weyl fermions, and we review the observation of high-fold chiral fermion phases by taking the rhodium silicide, RhSi, family of crystals as a prime example. Lastly, we discuss recent advances in Weyl-line phases in magnetic topological materials. With this Review, we aim to provide an introduction to the basic concepts underlying Weyl physics in condensed matter, and to representative materials and their electronic structures and topology as revealed by spectroscopic studies. We hope this work serves as a guide for future theoretical and experimental explorations of chiral fermions and related topological quantum systems with potentially enhanced functionalities.

\end{abstract}


\date{\today}

\maketitle

\textbf{Introduction}

The cross-pollination between high energy and condensed matter physics has led to a more profound understanding of fundamental organizing principles of matter such as the concepts of spontaneous symmetry breaking, phase transitions, and renormalization \cite{Weyl, Wilczek, Anderson}. Such knowledge has, in turn, greatly helped researchers understand inner workings of magnets, superconductors, and other exotic states of materials \cite{Thouless, Graphene, Hasan2010, Qi2011, TI_book_2014_2, Bansil_rev,Weyl_Vafek,Weng_rev, ARPESrev, Haldane_rev}. The observation of massless Dirac fermions in graphene and topological insulators has become a cornerstone of the past decade's research activity in electronic matter \cite{Graphene, Hasan2010, Qi2011, TI_book_2014_2, Bansil_rev}. In the past few years, Weyl and related materials are emerging as the new frontier along this line of research \cite{Weyl_Vafek, ARPESrev, Weyl2, Neumann, Volovik2003, herring_accidental_1937, abrikosov_properties_1971, nielsen1983adler, Murakami2007,Transportrev,BurkovRev,PuRev,PlankRev,STMrev,WeylRMP,TaAs_Hasan,TaAs_Ding,Huang2015,Weng2015,MIT_Weyl,Wan2011,Burkov2011,Balents_viewpoint,Transportrev2}. The realization of electronic Weyl fermions makes it possible to access a kaleidoscope of novel quantum phenomena in tunable systems \cite{ARPESrev, Weyl_Vafek, BurkovRev, Transportrev, PuRev, PlankRev, STMrev, WeylRMP,  TaAs_Hasan,Huang2015,Weng2015, TaAs_Ding,MIT_Weyl,Transportrev2}. Inspired by the richness of electronic structures in quantum matter, Weyl fermions were generalized to describe topological fermions beyond condensed matter analogs of the Standard Model of high energy physics \cite{WT-Weyl,ST-Weyl}. Such unconventional topological fermions form a catalog of zero-dimensional point degeneracies of bands in momentum space. These fermions can be classified by their topological invariants (Box 1), the dimensionality of momentum space in which they arise and the number of bands which cross to form the point degeneracy. Such unconventional fermions include higher-fold (three-fold, four-fold and six-fold) chiral fermions in three-dimensional crystals \cite{manes, New Fermion, KramersWeyl,RhSi,CoSi,FeSi, RhSi_exp,CoSi_exp,AlPt_exp,CoSi_exp2,FeSi_exp,acoustic1,acoustic2}. They are characterized by a Chern number topological invariant, also called the chiral charge \cite{ARPESrev, Weyl_Vafek}. The topological invariants associated with chiral fermions guarantee the existence of protected boundary modes, Fermi arcs, the open surface states on the terminated surface of the material \cite{Wan2011}. Recent theories found that all point-like degeneracies in non-magnetic structurally chiral crystals with spin-orbit coupling (SOC) are the chiral-charged fermions \cite{KramersWeyl}. In contrast, most high-fold degenerate fermions in structurally achiral space groups do not possess chiral charges, such as four-fold Dirac fermions \cite{3DDirac_theory, Na3Bi_theory,Na3Bi_exp_bulk,Hasan_Na3Bi} and eight-fold double Dirac fermions \cite{New Fermion, Double Dirac}. Another catalog of topological electronic structures consists of nodal lines, where bands are degenerate along an entire closed curve in momentum space. Such one-dimensional band crossings exhibit a winding number topological invariant which protects topological drumhead surface states (Box 1) \cite{Burkov2011_2, Phillips, Chiu, Chen_NL, Cu3XN_1,  Cu3XN_2, Ca3P2, PTS, ZrSiS_NC,ZrSiS,ZrSiSe_ZrSiTe,ZrSiSe_ZrSiTe_Neupane,ZrSiSe_Hao,Guang2}. Unlike point degeneracies, nodal lines can exhibit rich composite structures including nodal chains, Hopf links and nodal knots \cite{chain, Co2MnGa_theory, link2, link3, link4, link5,Co2MnGa_ARPES,ChangYee_WeylLink}. A third catalog consists of hybrid zero-dimensional and one-dimensional degeneracies, where three-fold point degeneracies serve as boundary points interconnecting two-fold nodal lines \cite{Nexus,trip1,trip2, extra_3fold, MoP,WC}. Topological fermions often give rise to anomalous electronic, magnetic, and optical response, especially in cases where the topology is associated with momentum-space singularities of Berry curvature (Box 1) \cite{ARPESrev, Weyl_Vafek, BurkovRev, Transportrev, PuRev, PlankRev, STMrev, WeylRMP,Transportrev2}. For example, topological fermions can support unconventional superconductivity \cite{Weyl-SC-1, Weyl-SC-2, Weyl-SC-5, Weyl-SC-7, sup_chiral1} or a large anomalous Hall effect in topological magnets \cite{AHE_Weyl}. Under an external magnetic field, Weyl fermions can realize the chiral anomaly \cite{Duval06, Son13,Burkov14,Ong_Chiral}, the chiral magnetic effect  \cite{Fukushima, gryo, Pesin}, the quantum nonlinear Hall effect \cite{Inti}, and nonlocal transport \cite{Nonlocal, Transmission}. The magnetic orbits arising from open Fermi arc surface states and bulk chiral Landau levels can induce unusual quantum oscillations \cite{Hosur, Ashvin_SdH,SdH_exp1,SdH_exp2}.  Weyl semimetals can also support exotic nonlinear optical response, which may be useful for new solar cells or light sensors \cite{PC_Weyl_Nagaosa,PC_Weyl_Moore,PC_Weyl_Chris,Moore2,TaAs_optics, arc_photo, TaAs_OP2, TaAs_exp2, MoTephoto_exp, CPGE_exp_Q}. The rich quantum properties of Weyl semimetals have been argued to have potential applications in the next-generation quantum devices \cite{ARPESrev, Weyl_Vafek, BurkovRev, Transportrev, PuRev, PlankRev, STMrev, WeylRMP,  TaAs_Hasan,Huang2015,Weng2015, TaAs_Ding,MIT_Weyl,Transportrev2}.

The goal of this article is to review the research progress of this developing field with a focus on electronic structure as viewed primarily by photoemission spectroscopy. Several excellent reviews have summarized the large body of works covering developments between 2015-18 \cite{Transportrev, PuRev, PlankRev, BurkovRev, STMrev,WeylRMP,Transportrev2}, hereafter reviewing the basics, we pay more attention to the latest progress and the new frontiers that emerged. Although the focus of this review is on spectroscopic experiments and materials development, we do include some basic theory in order to provide an overall conceptual scenario and experimental understanding of the materials. Section I starts with the basic theoretical background emphasizing the topological aspects. Section II reviews the experimental observation of the Weyl semimetal phase \cite{TaAs_Hasan, TaAs_Ding}, focusing on spectroscopic observation based on angle-resolved photoemission spectroscopy (ARPES) \cite{TaAs_Hasan, TaAs_Ding, NbAs_Hasan, TaAs_Ding_2, TaAs_Chen, TaP_Hasan, TaP_Shi, NbP_Hasan, TaP_Chen, spin1,spin2,NbP_STM1,TaAs_STM1,TaAs_STM2}. 
We then summarize experiments that report unusual aspects of magnetic and optical response in the TaAs family \cite{Chiral_anomaly_ChenGF, Chiral_anomaly_Jia, NbAs_Transport_1, NbAs_Transport_2, TaP_Transport_1, TaP_Transport_2, TaP_Chiral_anomaly_1, TaAs_optics, TaAs_exp2,NbP_MAG}. Section III highlights the experimental progress in type-II Weyl and related metals \cite{WT-Weyl,ST-Weyl, WMoTe-Weyl, MT-Weyl, MT-Weyl2, WT-ARPES-1, WT-ARPES-2, WT-ARPES-3, WT-ARPES-4, WT-ARPES-5,  WT-ARPES-8, WT-ARPES-9, WT-STM-1,WT-STM-2,LAG,tans1,tans2,monwte21,monwte22,monwte23,monwte24,RAlX,CeAlGe_transport}. Section IV introduces topological chiral crystals as generalized versions of Weyl semimetals \cite{KramersWeyl}, highlighting the recent observation of giant helicoid Fermi arcs in the RhSi family \cite{RhSi, CoSi,  RhSi_exp, CoSi_exp, AlPt_exp, CoSi_exp2}. Section V highlights the Weyl line phase with drumhead surface states observed in magnetic materials  \cite{Burkov2011_2, Phillips, Chiu, Chen_NL, Cu3XN_1,  Cu3XN_2, Ca3P2, PTS, ZrSiS_NC,ZrSiS,chain,Co2MnGa_theory, link2, link3, link4, link5,Co2MnGa_ARPES,ZrSiSe_ZrSiTe,ZrSiSe_ZrSiTe_Neupane,ZrSiSe_Hao,ChangYee_WeylLink,Guang2}. We conclude with an outlook which includes new research directions.

\bigskip

\textbf{I. Weyl fermions in condensed matter physics}

Weyl fermions arise as low-energy quasiparticle excitations in three-dimensional crystals lacking inversion symmetry or time-reversal symmetry or both \cite{ARPESrev, Weyl_Vafek, BurkovRev, Transportrev, PuRev, PlankRev, STMrev, WeylRMP,  TaAs_Hasan,Huang2015,Weng2015, TaAs_Ding,MIT_Weyl,Transportrev2}. In a Weyl semimetal, Weyl fermions are the sources/sinks of the Berry curvature field in momentum space (Box 1, Fig.~\ref{Fig1}\textbf{a}) \cite{herring_accidental_1937, abrikosov_properties_1971, nielsen1983adler, Murakami2007, Wan2011, Balents_viewpoint, Burkov2011,   Huang2015, Weng2015, TaAs_Hasan, MIT_Weyl, TaAs_Ding, ARPESrev, Weyl_Vafek, Transportrev, PuRev, PlankRev, STMrev, WeylRMP, Neumann, Volovik2003, BurkovRev,Transportrev2}. The integral of the Berry curvature field of any two-dimensional manifold enclosing a Weyl node is a quantized integer, which defines the Chern number/chiral charge of the Weyl fermion \cite{Wan2011}.  The bottom panel of Fig.~\ref{Fig1}\textbf{a} is a schematic of two oppositely-(chiral) charged Weyl fermions in bulk \cite{Wan2011}. A plane cutting between the Weyl fermions exhibits the Chern number $|C|=1$, whose projected Hamiltonian $H_{y}(k_{x}, k_{z})$  represents a two-dimensional Chern insulator with a one-dimensional chiral edge mode (Fig.~\ref{Fig1}\textbf{b}). Such a chiral edge state can be obtained from any plane between the two Weyl cones. In this way,  a topologically protected two-dimensional open surface state, the Fermi arc, is formed between two oppositely-charged Weyl fermions (Fig.~\ref{Fig1}\textbf{c}). The isoenergetic contour of a Fermi arc is an open Fermi surface connecting projected Weyl pockets (top panel, Fig.~\ref{Fig1}\textbf{a}). This is different from a conventional two-dimensional electron dispersion whose Fermi surface is a closed contour. What is more, Fermi arc surface states can take helicoidal structures described by what is mathematical known as non-compact Riemann surfaces \cite{Arc_Chen, RhSi_exp, Arc_photon}.

Weyl fermions are local singularities in energy and momentum, so their exotic response arises only in a finite topologically non-trivial energy and momentum window. In a simple picture, Weyl fermions can be viewed as arising from a band inversion, in which the conduction band sinks below the valence band in energy (Fig. \ref{Fig1}\textbf{d}) \cite{Wan2011, Balents_viewpoint, Burkov2011, Huang2015, Weng2015, MIT_Weyl}. In this picture, the topologically non-trivial window is closely related to the degree of band inversion, which determines the momentum-space separation of the Weyl fermions and their extent in energy \cite{WeylRMP, WMoTe-Weyl, LAG, NbP_Hasan}. For a given energy within the topological window, the electronic structure exhibits pockets associated with a non-zero chiral charge (top, Fig. \ref{Fig1}\textbf{e}). However, if one sweeps in energy out of the topological window, pockets with opposite chiral charge merge and are then associated with zero chiral charge (bottom, Fig. \ref{Fig1}\textbf{e}). Considering the surface states, within the topological energy window topological Fermi arcs connect bulk pockets with opposite chiral charge. Outside of the topological energy window, no Fermi arc is topologically protected, although studies do report that a non-topological arc may persist depending on details of the electronic structure of the specific material, the crystalline surface termination and other parameters (dashed line, Fig. \ref{Fig1}\textbf{e}). The topologically non-trivial window is an important consideration when exploring optical, magnetic or other responses of a material hosting Weyl fermions \cite{WeylRMP, WMoTe-Weyl, LAG, NbP_Hasan}. For example, for many applications it is crucial that the Fermi level sit within the topological window or that optical transitions occur within the topological window. The advent of Kramers-Weyl fermions in structurally chiral crystals, discussed in Sect. IV, marked an important development in the search for materials with larger topological windows. Chiral crystals naturally give rise to Weyl fermions pinned to Kramers points in the bulk Brillouin zone, producing Weyl fermions with maximal separation in momentum space (Fig. \ref{Fig1}\textbf{f}) \cite{KramersWeyl}. Such systems can similarly produce substantially larger topological energy windows than those known in achiral Weyl semimetals.

In the simplest case, a Weyl fermion is a two-fold degenerate band crossing of two singly-degenerate bands with linear dispersions in all three momentum space directions (Fig.~\ref{Fig1}\textbf{g}).  Due to fewer symmetry constraints, Weyl fermions in condensed matter physics have a higher degree of freedom and more affluent properties than relativistic Weyl fermions, such as type-II Weyl fermions (Fig.~\ref{Fig1}\textbf{h}) \cite{WT-Weyl, ST-Weyl}. The difference between type-I and type-II Weyl fermions is on the degree of Weyl cone tilt. The two bands of type-I Weyl fermions though can also be slightly-tilted but still keep the opposite signs of Fermi velocities in all directions. In contrast, the bands forming type-II Weyl fermions are strongly tilted at least in one direction where they have the same sign of Fermi velocities \cite{WT-Weyl, ST-Weyl}.

Under additional crystal symmetries, Weyl fermions can have other types of electronic structures, rather than the conventional two-fold degenerate linear dispersions. For example, four-fold or six-fold rotation axes allow double-Weyl fermions with charge $|C|=2$. Six-fold rotation axes allow triple-Weyl  fermions  with charge $|C|=3$ \cite{double}. The energy dispersion of a double-Weyl (triple-Weyl) fermion is still linear along the rotation axis but quadratic (cubic)  in the plane normal to the axis (Fig.~\ref{Fig1}\textbf{i}). A double-Weyl (triple-Weyl) fermion can be viewed as two (three) singly-charged Weyl fermions stick together in momentum-space.  If the high-fold rotation symmetry is broken, a double-Weyl (triple-Weyl) fermion will split into two (three) conventional singly-charged Weyl fermions \cite{double, HgCrSe, SrSi2}.

In non-magnetic three-dimensional crystals, apart from  two-fold degenerate Weyl fermions,  point fermions with three-, four-, six-, and eight-fold degeneracy are allowed \cite{manes, New Fermion,KramersWeyl,RhSi,CoSi,RhSi_exp,CoSi_exp2,CoSi_exp,AlPt_exp,FeSi,FeSi_exp,Double Dirac,trip1,trip2,Nexus,extra_3fold,MoP,WC,acoustic1,acoustic2}. These high-fold fermions can be classified into two groups: chiral fermions and non-chiral fermions. High-fold chiral fermions are natural generalizations of two-fold Weyl fermions  \cite{KramersWeyl, manes,New Fermion, RhSi, CoSi}.  There are two approaches to classify the topology of the $n$-fold chiral fermions. One way is to determine the Chern number of each band \cite{New Fermion}, and the other way is to calculate the chiral charges for the $n-1$ band gaps [$C_{1}$, $C_{2}$,..., $C_{n-1}$] (Fig.~\ref{Fig1}\textbf{j}) \cite{RhSi}.  The other group of high-fold fermions is non-chiral which has no well-defined chiral charge \cite{Double Dirac,trip1,trip2,Nexus,extra_3fold,MoP,WC}. In most cases, one can not determine whether a $n$-fold fermion is chiral or not simply from the point node degeneracy. High-fold degenerate fermions display complex space groups dependent on band dispersions and topology. For example, the three-fold fermions at the $P$-point in the space group No. 199 are the spin-1 generalization of Weyl fermions \cite{New Fermion},  while the three-fold fermions in the space group No. \#220 and the three-fold linking points in the space groups Nos. 156-159 and Nos.  187-189 do not have quantized chiral charges \cite{trip1,trip2,Nexus,extra_3fold,MoP,WC}. Four-fold Dirac fermions in centrosymmetric crystals are non-chiral \cite{WeylRMP}, while four-fold fermions at the $\Gamma$ point of cubic chiral crystals have quantized chiral charges  \cite{KramersWeyl, RhSi, CoSi}. It has been found that all point degeneracies in non-magnetic chiral crystals are chiral fermions with quantized non-zero Chern numbers \cite{KramersWeyl}. This universal topological electronic property of all non-magnetic chiral crystals provides a powerful tool for searching for new topological chiral materials.

\bigskip

\textbf{II. First Weyl semimetals: the TaAs family}

The material realization of Weyl topology was inspired by the development of topological insulators. The first materials prediction was in the pyrochlore iridates family, R$_2$Ir$_2$O$_7$ (R is a rare earth element) by Wan \textit{et al.} in 2011 \cite{Wan2011}. To realize Weyl fermions, the authors argued for a special kind of anti-ferromagnetic order with an all-in/all-out configuration to break time-reversal symmetry of R$_2$Ir$_2$O$_7$ \cite{Wan2011}.   ARPES measurements on these materials have been lacking. Moreover, the assumed magnetic order is still under debate in experiments \cite{Iridate_Mag_1, Iridate_Mag_2}. Following their footsteps, another work proposed to realize Weyl fermions in a superlattice consisting of a stack of alternating thin films of a topological insulator and a normal ferromagnetic insulator \cite{Burkov2011}. No experiment has been reported, possibly owing to the lack of a suitable ferromagnetic insulator whose lattice matches with known topological insulators. The next proposal made was HgCr$_2$Se$_4$ \cite{HgCrSe}, which is known to be a natural ferromagnet with a fairly high Curie temperature $T_{\textrm{Curie}}\simeq120$ K  \cite{HgCrSe_Transport}.  However, ARPES experiments have not been successful on this compound.

The breakthrough finally happened in early 2015 as the first Weyl fermion material was predicted and experimentally confirmed in the inversion symmetry breaking crystal TaAs family \cite{Weng2015, Huang2015, TaAs_Hasan, TaAs_Ding}. The TaAs class of crystals have a body-centered tetragonal lattice system in the space group $I4_1md$ ($\#109$). The crystal consists of interpenetrating Ta/Nb and As/P sub-lattices (Fig.~\ref{Fig2}\textbf{a}). The lattice lacks inversion symmetry, which is the key condition for realizing Weyl fermions. According to density functional theory (DFT) calculations, there are two groups of Weyl fermions near the Fermi level in TaAs and its cousins. One group of  Weyl fermions $W_{1}$ lies at  $k_{z}=2\pi/c$ (cyan plane Fig.~\ref{Fig2}\textbf{b}), and the other group $W_{2}$ lies closer to $\Gamma$ \cite{Weng2015, Huang2015}.  The  $W_{2}$ Weyl fermions and their Fermi arc surface states have been directly observed by  ARPES \cite{TaAs_Hasan, TaAs_Ding,  TaAs_Chen}. Figure \ref{Fig2}\textbf{c} shows the Fermi surface obtained by  the soft X-ray (SX) ARPES, which selectively enhances the states from the bulk at the Fermi level  by choosing a higher incident photon energy \cite{TaAs_Hasan}. On the isoenergetic contour, ARPES observed eight isolated Fermi contours (indicated by the arrows in Fig. \ref{Fig2}\textbf{c}). The $k$-space locations and separations of the bulk Fermi points are consistent with the location of  $W_{2}$ Weyl fermions in DFT calculations \cite{TaAs_Hasan, TaAs_Ding,  TaAs_Chen}. Moreover, the bands disperse linearly away from the gapless nodes, which further indicates that they are Weyl fermions (Fig.~\ref{Fig2}\textbf{d}). To provide further evidence for Weyl fermions, the experimental observation of Fermi arc surface states was needed. On the (001) surface, two $W_{2}$ Weyl fermions of the same chiral charge project at the same $k$-point. From the topological bulk-boundary correspondence, there should therefore be two Fermi arc surface states originating from one $k$-point on the surface. Indeed, on the surface, ultraviolet (UV) ARPES observed crescent-shaped surface states, which are the Fermi arcs according to theoretical predictions (Fig.~\ref{Fig2}\textbf{e}). When the SX-ARPES-measured bulk nodes are superimposed onto the low-photon-energy UV-ARPES-measured Fermi surfaces, the surface Fermi arc terminations match with the projection of the bulk Weyl fermions on the surface  Brillouin zone (BZ) (Figs.~\ref{Fig2}\textbf{f,g}) \cite{TaAs_Hasan}. The bulk and surface evidence, taken collectively, revealed the Weyl fermions in TaAs. Similar surface states were observed in the other materials of the TaAs family: NbAs, TaP, and NbP \cite{NbAs_Hasan, TaAs_Ding_2, TaAs_Chen, TaP_Hasan, TaP_Shi, NbP_Hasan, TaP_Chen}. A large spin polarization of Fermi arcs, approaching 80\%, was observed in TaAs. The measured spin texture of the Fermi arcs is consistent with theoretical predictions \cite{spin1,spin2}. Scanning tunneling microscopy (STM) measurements have also observed the quasiparticle interference (QPI) of the TaAs family, which also observed the signature of Fermi arcs \cite{NbP_STM1, TaAs_STM1, TaAs_STM2}. The complementary spectroscopy measurements in different compounds from different theoretical and experimental groups provide further evidence of Weyl fermions in the TaAs family \cite{Weng2015, Huang2015, TaAs_Hasan, TaAs_Ding, NbAs_Hasan, TaAs_Ding_2, TaAs_Chen, TaP_Hasan, TaP_Shi, NbP_Hasan, TaP_Chen,  NbP_STM1, TaAs_STM1, TaAs_STM2,spin1,spin2}. Apart from electronic systems,  Weyl excitations have also been observed in the gyroid photonic crystal, around the same time \cite{MIT_Weyl}.

Weyl semimetals exhibit several exotic magnetic and optical responses. One of the most fundamental examples is the chiral anomaly \cite{nielsen1983adler, ABJ1, ABJ2}. Historically, the chiral anomaly was crucial in understanding many important aspects of the Standard Model in particle physics. The best-known case is the triangle anomaly associated with the decay of the neutral pion $\pi^{0}$ \cite{ABJ1, ABJ2}. This chiral magnetoresistance effect can be realized in Weyl semimetals \cite{Ong_Chiral}. For instance, under parallel magnetic and electric fields ($\vec E  ||  \vec B$), electrons from one Weyl cone are pumped to other Weyl cone of the opposite chirality, causing unbalanced charge distributions between the two Weyl cones (Fig.~\ref{Fig2}\textbf{h}). The chiral anomaly is associated with an axial charge current leading to the negative longitudinal magnetoresistance in Weyl semimetals, where the conductivity increases under an increasing external magnetic field. For example, the negative magnetoresistance in the Dirac semimetal Na$_{3}$Bi is a strong hint of the existing of Weyl fermions under an external magnetic field \cite{Ong_Chiral}. The observation of Weyl fermions in TaAs provides a natural route to realizing the chiral anomaly in condensed matter.  A strong negative magnetoresistance has been experimentally observed in the TaAs family, which was interpreted as a signature of the chiral anomaly \cite{Chiral_anomaly_ChenGF, Chiral_anomaly_Jia, NbAs_Transport_1,  TaP_Transport_2, TaP_Chiral_anomaly_1}.  However, some experimental and theoretical work has also found that negative longitudinal magnetoresistance can also even be realized in crystals without Weyl fermions \cite{Effect1, Effect3,Effect4}.  Since multiple effects can give rise to the negative magnetoresistance, decisive experimental evidence of the chiral anomaly in Weyl semimetals has yet to emerge.  Large positive magnetoresistance is another signature/hint of Weyl fermions \cite{Cd3As2_MAG, Cd3As2_MAG_theory}, which has also been reported in NbP \cite{NbP_MAG, TaP_Transport_1}. It also needs to be noted that some other effects, such as balanced electron and holes carriers, can also induce colossal magnetoresistance  \cite{NbAs_Transport_2, WP2_ARPES}.

Another important technique to detect the Berry curvature field of Weyl fermions is via their nonlinear optical responses. One example is the circular photogalvanic effect (CPGE), an optical response associated with photocurrents induced by circularly-polarized light \cite{PC_Weyl_Chris, TaAs_optics}. The asymmetric particle-hole excitation of a Weyl fermion creates a chirality-dependent CPGE photocurrent (Fig.~\ref{Fig2}\textbf{i}) \cite{PC_Weyl_Chris}. A circularly polarized photon excites only one side of the Weyl cone and generates a photocurrent that is proportional to its chiral charge. Weyl fermions of opposite chiral charges generate a photocurrent propagating in opposite directions.  The sum of photocurrents from a relativistic Weyl fermion pair must vanish identically. Due to the small tilt of the Weyl fermions,  the net photocurrent in TaAs does not vanish \cite{PC_Weyl_Chris, TaAs_optics}. It is important to note that the CPGE can also arise in the absence of Weyl fermions. To allow direct transitions within the Weyl cones, while blocking transitions to/from other bands, a laser of low photon energy has been necessary in experiments. The measured CPGE under mid-infrared lasers ($\hbar \omega = 120$ meV) oscillates with changing laser polarization (Fig.~\ref{Fig2}\textbf{i}). The CPGE photocurrent has maximum value for right-handed (RCP) light, minimum value for left-handed (LCP) laser, and zero for linearly polarized light. These observations are consistent with the chirality of the Weyl fermions in TaAs \cite{TaAs_optics}. Recent experiments have further observed shift currents dominated by the bulk Weyl fermions in TaAs \cite{TaAs_exp2}. The giant anisotropic second-harmonic generation has also been observed in TaAs \cite{TaAs_OP2}.

To conclude this section, we discuss how to determine the Weyl fermion chiral charge by counting Fermi arcs in ARPES (Fig. 3). As two contrasting examples, we compare the Fermi arc surface states in the Weyl semimetal TaAs and the Dirac semimetal Na$_{3}$Bi \cite{3DDirac_theory, Na3Bi_theory, Na3Bi_exp_bulk, Hasan_Na3Bi}. A symmetry-protected three-dimensional Dirac fermion is a four-fold degeneracy that can be viewed as a composition of two oppositely-charged Weyl fermions \cite{3DDirac_theory}. As a result, each Dirac point is expected to be connected by two Fermi arcs on the surface \cite{Hasan_Na3Bi}. However, the two Fermi arcs of a Dirac fermion ($C=0$) is entirely different from the case where two Fermi arcs are induced by the chiral charge $|C|=2$. On the  (100) surface of Na$_{3}$Bi, two Fermi arcs (the white dotted lines) connecting the Dirac points (the black dots) form a closed loop (Fig. 3\textbf{a, b}). Along with a cut passing between the Dirac nodes (path $\beta$,  Fig. 3\textbf{c}), there are two surface chiral modes (white dashed lines)  with opposite signs of Fermi velocity. Thus the net chirality along the path is zero, consistent with the chiral charge of a Dirac fermion, $C=0$. Now we check the chiral modes of the Weyl semimetal TaAs family.  Two chiral modes with the same sign of Fermi velocity are observed along the loop $p$ enclosing the projected Weyl fermions (Fig. 3\textbf{d}). This is consistent with the net chiral charges $|C|=2$ enclosed in the loop. Counting the net chiral surface modes in ARPES is a powerful tool to determine the absolute value of the net Chern number. This method fails in ARPES where bulk pockets are crossing the the Fermi level. For example, the net chiral modes along the triangle path $c$ is two, inconsistent with the net charge $|C|=3$ inside (Figs. 3\textbf{d,f}). Clear bulk states can be observed near $\bar{X}$ along the path. In this case, without accessing states above the Fermi level,  one can not count the chiral edge modes along the entire loop by ARPES. Therefore, to determine the Chern number in ARPES,  the energy dispersion must be fully gapped in bulk everywhere along the chosen momentum-space path.

\bigskip

\textbf{III. Type-II Weyl semimetal: the LaAlGe family and the Mo$_{x}$W$_{1-x}$Te$_{2}$ family}

Shortly after the observation of Weyl fermions in the TaAs family, a distinct kind of Weyl fermion was theoretically proposed, termed the type-II Weyl fermion, in contrast to the Weyl fermions of TaAs, which were then referred to as type-I  \cite{WT-Weyl, ST-Weyl}. The two bands forming type-II Weyl fermions have the same sign of Fermi velocity along a specific path in momentum space (Fig.~\ref{Fig4}\textbf{a}). While type-I Weyl fermions have point Fermi surfaces (Fig.~\ref{Fig1}\textbf{e}), type-II Weyl fermions have electron and hole pockets touching at a point (top panel of Fig.~\ref{Fig4}\textbf{b}). At an energy slightly off the type-II Weyl node, the electron and hole pockets disconnect, and the type-II Weyl fermion is enclosed by one of the two pockets (bottom panel of Fig.~\ref{Fig4}\textbf{b}).

Type-II Weyl semimetal was first predicted in the Mo$_{x}$W$_{1-x}$Te$_{2}$ family, space group \#31 \cite{WT-Weyl, WMoTe-Weyl, MT-Weyl, MT-Weyl2}. At the Fermi level of  the Mo$_{x}$W$_{1-x}$Te$_{2}$ family, a large arc-like feature and two large bulk projections have been observed by ARPES (Fig.~\ref{Fig4}\textbf{c})  \cite{WT-ARPES-1, WT-ARPES-2, WT-ARPES-3, WT-ARPES-4, WT-ARPES-5,  WT-ARPES-8, WT-ARPES-9, WT-STM-1,WT-STM-2}. This surface arc was interpreted as the topological Fermi arc due to Weyl fermions. Later research found that the arc-like surface state is topologically trivial and can exist even without Weyl fermions \cite{WT-ARPES-5, LAG,WT-STM-1,WT-ARPES-8, WT-ARPES-9,WT-STM-2}. Some experiments attempted to seek other tiny arc-like features \cite{WT-ARPES-8, WT-ARPES-9} or to resolve the states above the Fermi level by pump-probe ARPES (Fig. \ref{Fig4}\textbf{d}) and STM \cite{WT-ARPES-3,WT-STM-1,WT-STM-2}.  These searches for the Weyl semimetal phases in the Mo$_{x}$W$_{1-x}$Te$_{2}$ family typically relied heavily on the comparison between experimental measurements and DFT calculations. According to DFT, the type-II Weyl fermions are around 50 meV above the Fermi level, and the band inversion is of the order of 10 meV.  At the Fermi level, the oppositely-charged Weyl fermions merge into a trivial pocket. As a result, the Weyl semimetal phase in  Mo$_{x}$W$_{1-x}$Te$_{2}$ is near a topological phase transition in first-principles calculations. For example, under slight differences of lattice constants, WTe$_{2}$ was proposed to be either a trivial insulator \cite{WMoTe-Weyl} or a Weyl semimetal \cite{WT-Weyl}. Similarly, depending on the size of the unit cell, MoTe$_{2}$ was predicted to contain either 8 Weyl fermions \cite{MT-Weyl} or 4 Weyl fermions plus extra nodal lines \cite{MT-Weyl2}. However, these different topological phases were associated with small differences in electronic structure, which were found to be beyond the experimental resolution of ARPES measurements. Therefore, the ``overall agreements" between theoretical calculations and experimental measurements are insufficient to determine the Weyl phase. As a result, despite the intensive experimental studies, the decisive evidence for the type-II Weyl fermions in Mo$_{x}$W$_{1-x}$Te$_{2}$ is still needed \cite{WT-ARPES-1, WT-ARPES-2, WT-ARPES-3, WT-ARPES-4, WT-ARPES-5,  WT-ARPES-8, WT-ARPES-9, WT-STM-1,WT-STM-2}. Beyond the Weyl physics, the layered transition-metal dichalcogenide WTe$_{2}$ and MoTe$_{2}$ have many other interesting properties \cite{tans1, tans2, monwte21, monwte22, monwte23, monwte24}. For example, large non-saturating magnetoresistance has been observed in WTe$_{2}$ \cite{tans1, tans2}. The monolayer WTe$_{2}$ realizes two-dimensional quantum spin Hall effect up to 100 kelvin \cite{monwte21, monwte22, monwte23, monwte24}.

More examples of type-II Weyl semimetals were found in the LaAlGe family in the inversion-breaking space group \#109 \cite{LAG}. SX-ARPES observed two bulk Fermi surfaces connected at two isolated points at the Fermi level of LaAlGe (the left panel of Fig.~\ref{Fig4}\textbf{e}). At binding energy 0.25 ~eV below the Fermi level, the two Fermi pockets disconnect (right panel, Fig.~\ref{Fig4}\textbf{e}). It was further observed that the upper (lower) pocket is the hole (electron) pocket since it expands (contracts) with deeper binding energy. This behavior is consistent with the bulk Fermi surfaces of type-II Weyl fermions (Fig.~\ref{Fig4}\textbf{b}).  The observed node arises from the crossing between two bands with linear dispersions along with all three momentum-space directions. Along the $k_{y}$ direction, the Fermi velocities of the two bands have the same sign of Fermi velocity (Fig. \ref{Fig4}\textbf{f}), again consistent with the electronic structure of type-II Weyl fermions (Fig.~\ref{Fig4}\textbf{a}). Based on the SX-ARPES data, these observed band crossings are located at a generic momentum-point. These systematic data demonstrate that  LaAlGe hosts type-II Weyl fermions. Due to hybridization between the surface Fermi arcs and trivial bulk pockets, the Fermi arcs in LaAlGe were found to have weak intensity even in DFT calculations  and have yet to be observed in experiments. 
The other two compounds in the LaAlGe family are the noncentrosymmetric magnets CeAlGe and PrAlGe \cite{RAlX}. In this sense, this class of materials covers many varieties of Weyl semimetals, including type-I, type-II, inversion breaking, and time-reversal breaking, depending on a suitable choice of the rare-Earth element. Recent experiments further observed the singular angular magnetoresistance in CeAlGe, proposed to exhibit a switching mechanism that flips the electrical resistance for specific field orientations, which can be used to engineer a magnetic switch with magnetic-field-orientation-driven on/off states \cite{CeAlGe_transport}.

\bigskip

\textbf{IV. Topological chiral crystals in the RhSi family}

Despite a growing list of experimentally-observed Weyl semimetals and candidates developed in 2015 and the following years, there remained an intense interest in developing materials with simpler electronic structures  \cite{Transportrev,BurkovRev,PuRev,PlankRev,STMrev,WeylRMP,Transportrev2}. With a focus on Weyl semimetals brought about by band inversion, first-principles calculations of a wide range of crystals found that in most materials (1) the Weyl fermions are too close together in momentum space, typically leading to (2) short Fermi arcs and (3) narrow topologically non-trivial energy windows. Moreover predicted band structures often exhibit (4) a very large number of Weyl fermions scattered throughout the bulk Brillouin zone with (5) Weyl fermions that lie far from the Fermi level as well as (6) irrelevant trivial electron bulk and surface states at the Fermi level.
 
Key insights arose with a renewed focus on point touchings of bands at high-symmetry points in the Brillouin zone, which began to move the focus beyond a band-inversion picture. In principle, such symmetry-protected band crossings had previously been explored \cite{WeylRMP}. One point of view considered how high-fold degenerate band touchings in non-symmorphic space groups at high-symmetry points could be viewed as ``unconventional fermions'' beyond the Weyl, Dirac and Majorana fermions well-known from quantum field theory \cite{New Fermion}. Some of these symmetry-protected high-fold degenerate fermions at high-symmetry points were found to be Weyl-like in that they exhibit a chiral charge. Soon thereafter, it was discovered that for symmorphic chiral space groups time-reversal symmetry alone is sufficient to produce a Weyl point and that in fact \textit{all} Kramers points in these space groups pin Weyl points \cite{KramersWeyl}. For non-symmorphic chiral space groups with time-reversal symmetry, the $\Gamma$ point supports Kramers-Weyl fermions, while the Brillouin zone boundary exhibits either Kramers-Weyl fermions or chirally-charged nodal surfaces, depending on the specific space group (Figs.~\ref{Fig1}\textbf{f}-\textbf{i}). Meanwhile, the cubic chiral space groups Nos. 195-199 and Nos. 207-214, which have additional symmetries, allow three-, four-, and six-fold chiral fermions (Fig.~\ref{Fig1}\textbf{j}) \cite{New Fermion, KramersWeyl}. This understanding gradually eroded the prevailing intuition that Chern numbers in three-dimensional crystals were necessarily associated with accidental band degeneracies at generic points in momentum space. Instead, Weyl fermions and their higher-fold chirally-charged cousins were shown to exist universally in chiral crystal structures, and to be pinned to high-symmetry momenta. This understanding further brought to light the natural connection between the common notion of spatial chirality, or handedness, and the momentum-space chirality of a Weyl fermion \cite{KramersWeyl}. Theoretical studies have also found the topological chiral crystals can exhibit many exotic electronic, optical, and magnetic properties (Fig.~\ref{Fig5}\textbf{a}).

The development of topological chiral crystals immediately led to the groundbreaking prediction of a near-ideal Weyl semimetal in the RhSi family of materials \cite{RhSi, CoSi}. RhSi crystallizes in cubic non-symmorphic chiral space group No. 198. Its electronic structure was predicted to be dominated by two higher-fold chiral fermions near the Fermi level, one at the $\Gamma$ point and its oppositely-charged partner at the $R$ point (corner of the bulk Brillouin zone, Figs.~\ref{Fig5}\textbf{b, c}), representing the largest momentum-space separation possible between two chiral charges in the Brillouin zone. Calculations of the (001) surface electronic structure further predicted giant topological Fermi arcs stretching across the entire surface Brillouin zone, from $\bar{\Gamma}$ to the corner, $\bar{M}$ (Fig.~\ref{Fig5}\textbf{d}, inset). Subsequent experiments by photoemission spectroscopy explored the topological electronic structure in RhSi as well as its isostructural cousins CoSi and AlPt \cite{RhSi_exp, CoSi_exp, AlPt_exp, CoSi_exp2}. First, experiments by surface-sensitive UV-ARPES demonstrated topological Fermi arcs by directly counting chiral edge modes on momentum-space loop cuts through the surface Brillouin zone, obtaining a Chern number of $\pm 2$ (Fig.~\ref{Fig5}\textbf{e}) \cite{RhSi_exp, CoSi_exp, AlPt_exp}. These experiments also directly observed the helicoid structure of the topological surface states, with the Fermi arcs spiraling around $\bar{M}$ with varying binding energy (Fig.~\ref{Fig5}\textbf{f}). Lastly, the dispersion of the Fermi arcs was found to be consistent with first-principles calculations. Complementing these surface-sensitive results, bulk-sensitive SX-ARPES experiments observed a linear electronic dispersion around the $R$ point (Fig.~\ref{Fig5}\textbf{c}), consistent with the high-fold chiral fermions predicted in calculation. The discovery of a Weyl semimetal in the RhSi family marked a considerable advance over the first generation of Weyl semimetals. For example, the momentum separation of the oppositely-charged high-fold chiral fermions in RhSi is $\Delta = 1.16\ {\rm \AA}^{-1}$, an order of magnitude larger than the separation of Weyl point partners in TaAs, $\Delta = 0.07\ {\rm \AA}^{-1}$ \cite{RhSi, CoSi, Weng2015, Huang2015}. Naturally, the associated topological Fermi arcs observed in the RhSi family are also substantially longer than those observed in TaAs. 

Topological chiral crystals are predicted to exhibit related exotic phenomena (Fig.~\ref{Fig5}\textbf{a}), notably the quantized circular photogalvanic effect (CPGE) \cite{KramersWeyl}. In the quantized CPGE, photocurrents produced by circularly polarized light are closely related to a universal quantity written in terms of only fundamental constants and the topological chiral charge \cite{KramersWeyl, RhSi, Moore2, CPGE_exp_Q}. The quantized CPGE was first proposed in conventional two-fold Weyl semimetals \cite{Moore2}. Later theoretical work found that the quantized CPGE may also arise from higher-fold chiral fermions \cite{RhSi}. The theoretically simulated quantized CPGE from a two-fold Weyl fermion and four-fold chiral fermions tuned to half-filling are shown in Fig.~\ref{Fig5}\textbf{h}. The time derivative of the injection current is related to the Chern number in the gap. Crucially, for this effect chiral fermions of opposite chiral charge must sit at different energies in the electronic structure. This requirement excludes achiral Weyl semimetals, such as the TaAs family, where the presence of mirror or other roto-inversion symmetries pairs all $+$ Weyl fermions with $-$ partners at the same energy. By contrast, RhSi hosts oppositely-charged high-fold chiral fermions with a large energy difference of $\sim 0.4$ eV. Furthermore, RhSi has only two chiral fermions and few irrelevant electronic states near the Fermi level, which is expected to simplify experimental studies (Fig.~\ref{Fig5}\textbf{i}) \cite{RhSi}. Indeed, signatures of a topological photocurrent in RhSi have already been reported at room temperature \cite{CPGE_exp_Q}. The chiral magnetic effect similarly requires a relative energy offset in the Weyl fermions and may be explored in RhSi \cite{Fukushima, gryo, Pesin}.

First-principles calculations predict analogous higher-fold chiral fermions in other nonmagnetic members of the RhSi family, including RhGe, CoGe, AlPd and AuBe, as well as the closely-related BaPt$X$ series, $X = $ P, As, Sb \cite{RhSi,AlPt_two}. Moreover, while recent ARPES experiments have demonstrated a chiral charge in the RhSi materials, direct spectroscopic observation of the complete topological index remains missing. To achieve this, direct observation of the $n$ branches of the higher-fold chiral fermion is needed, along with the corresponding $n-1$ sets of Fermi arcs within each band gap \cite{RhSi_two}.   Analogously, tuning the pump laser wavelength or sample Fermi level in a quantized CPGE experiment may also allow the observation of different quantized CPGE values associated with the distinct chiral charges of the $n-1$ band gaps \cite{RhSi}. In neighboring communities, near-ideal Weyl phases have also been explored in the acoustic and optical phonon spectra of chiral crystals (Fig. ~\ref{Fig5}\textbf{i}) \cite{acoustic1, acoustic2, FeSi, FeSi_exp}. Negative refraction of surface acoustic waves has also been experimentally realized in these chiral metamaterials with giant Fermi arcs \cite{acoustic2}. It should similarly be possible to realize negative refraction of electrons in chiral crystals using the large Fermi arcs. The large number of candidates and rich topological structure of the RhSi family offers a rich playground for future work.

\bigskip

\textbf{V. Topological nodal line semimetals}

Weyl fermions or high-fold chiral fermions discussed above are zero-dimensional topological singularities in momentum space. In condensed matter physics, topological singularities can also be higher-dimensional. For instance, one-dimensional topological nodal lines can arise, where the conduction and valence bands cross each other along curves in momentum space, rather than at discrete points (Fig.~\ref{Fig6}\textbf{a}) \cite{Burkov2011_2, Phillips, Chiu, Chen_NL, Cu3XN_1,  Cu3XN_2, Ca3P2, PTS, ZrSiS_NC,ZrSiS,ZrSiSe_ZrSiTe,ZrSiSe_ZrSiTe_Neupane,ZrSiSe_Hao,Guang2}. The appropriate topological index for a nodal line is the winding number, given by the integral of the Berry connection along a closed loop encircling the nodal line  (Box 1).  The nontrivial topology of the nodal line protects  drumhead surface states connecting the bulk line nodes (Fig.~\ref{Fig7}\textbf{a})  \cite{Burkov2011_2, Phillips, Chiu, Chen_NL, Cu3XN_1, Cu3XN_2, Ca3P2, PTS, ZrSiS_NC,ZrSiS, ZrSiSe_ZrSiTe,ZrSiSe_ZrSiTe_Neupane,ZrSiSe_Hao,Guang2}. Nodal line fermions in momentum space are typically protected by certain crystal symmetries. For example, time-reversal and inversion symmetry together can protect nodal lines in the absence of SOC \cite{Chiu, Chen_NL,Cu3XN_1,Cu3XN_2}.  Mirror symmetry can also protect nodal lines, both in the presence and absence of SOC, when the band have opposite mirror eigenvalues  \cite{Chiu, Chen_NL,Ca3P2, PTS, ZrSiS_NC,ZrSiS,ZrSiSe_ZrSiTe,ZrSiSe_ZrSiTe_Neupane,ZrSiSe_Hao}. When time-reversal symmetry or inversion symmetry are broken, the nodal lines are two-fold degenerate and thus be called as Weyl lines \cite{PTS}.

Nodal lines and drumhead surface states were first proposed in 2011 \cite{Burkov2011_2}. Since then, many materials have been predicted to be nodal line semimetals, such as Cu$_{3}$NZn \cite{Cu3XN_1}, Cu$_{3}$NPd \cite{Cu3XN_2}, Ca$_{3}$P$_{2}$ \cite{Ca3P2}, and TlTaSe$_{2}$ \cite{Guang2}. Most of these predicted nodal lines in earlier studies were stabilized in the absence of SOC. After turning on SOC, these nodal lines will usually be gapped out. PbTaSe$_{2}$ is a rare case where Weyl lines can remain gapless after the inclusion of SOC when the opposite mirror eigenvalues are still preserved \cite{PTS}. The band crossings of the Weyl lines along the high-symmetry lines are indicated in the red box of Fig.~\ref{Fig6}\textbf{b}. The potential drumhead surface states of PbTaSe$_{2}$ were observed by ARPES (Fig.~\ref{Fig6}\textbf{c}), in good agreement with first-principles calculations. Nodal lines have also been observed in ZrSi$X$ ($X = $ S, Se, Te) \cite{ZrSiS_NC, ZrSiS}. According to DFT calculations, in the absence of SOC, the nodal lines in ZrSiS are protected by non-symmorphic mirror planes (top panel of Fig.~\ref{Fig6}\textbf{d}). In the presence of SOC, these nodal lines will hybridize and open band gaps (bottom panel of Fig.~\ref{Fig6}\textbf{d}). Several groups have observed clear diamond-shaped nodal lines with sharp linear dispersions in ZrSi$X$ (Figs.~\ref{Fig6}\textbf{e,f}) \cite{ZrSiS_NC, ZrSiS,ZrSiSe_ZrSiTe,ZrSiSe_ZrSiTe_Neupane,ZrSiSe_Hao}.

Recent theory has discovered systems hosting networks consisting of multiple nodal lines \cite{chain, Co2MnGa_theory, link2, link3, link4, link5,ChangYee_WeylLink}. The topological nodal chain consists of two nodal lines touching at a two-fold degenerate point (red and blue rings, Fig.~\ref{Fig7}\textbf{b}) was first proposed in crystals with non-symmorphic glide symmetries \cite{chain}. Later work found that the Weyl/nodal chains can also be protected by symmorphic mirror planes \cite{Co2MnGa_theory}. Another unconventional three-dimensional nodal-network is the Hopf-link where two nodal line link each other (red and yellow rings, Fig.~\ref{Fig7}\textbf{b}) \cite{Co2MnGa_theory,link2, link3,ChangYee_WeylLink}. The nodal knot, where a single nodal line entangled itself, has also been theoretically proposed in toy models \cite{link4, link5}. 

Despite the quick development of the theory of nodal line networks, their experimental realization has been limited by the availability of material candidates. Up to now, the topological nodal chain has only been experimentally realized in photonic crystals \cite{nodalchian_photon}. The other material candidate hosting Weyl line networks with the nodal-chain and Hopf-like link structures is the ferromagnetic metal Co$_{2}$MnGa family \cite{Co2MnGa_ARPES}. Three types of Weyl lines are predicted in Co$_{2}$MnGa, which are colored by blue, red, and yellow in Fig.~\ref{Fig7}\textbf{c}, respectively. Quite generally, after taking into account SOC, the only mirror plane which is preserved is the one whose normal is parallel to the direction of magnetization. The Weyl lines on the other mirror planes will open tiny band gaps ($\sim$ 1 meV). Recently,  both bulk Weyl lines and drumhead surface states were observed in Co$_{2}$MnGa \cite{Co2MnGa_ARPES,Comment_mag}. ARPES-measured isoenergetic contours reveal the signatures of three kinds of Weyl lines in Co$_{2}$MnGa (Fig.~\ref{Fig7}\textbf{d}).  Figure ~\ref{Fig7}\textbf{e} shows the energy dispersions of the blue line nodes along $k_{x}$ direction at different $k_{y}$ values. The fact that these crossings persist in a range of $k_{y}$ and move in energy is consistent with the interpretation of a Weyl line. The drumhead surface states connecting the yellow Weyl lines were observed in photoemission spectra acquired along the $k_{xy}$ direction. As shown in Fig.~\ref{Fig7}\textbf{f}, the bulk cones are indicated by the yellow dots, and the drumhead surface states are highlighted by the green dots. 

Recent work suggests that Weyl lines are exceptionally effective at producing large and robust anomalous Hall and Nernst effects (AHE, ANE). Co$_{2}$MnGa was observed to exhibit the largest ANE known at room temperature and exceeding other magnetic conductors by one order of magnitude \cite{Co2MnGa_TP1, Co2MnGa_TP2}. The AHE in Co$_{2}$MnGa, of order $10^3\ \Omega^{-1} \textrm{cm}^{-1}$, is also among the largest known in all materials and persists to room temperature \cite{Co2MnGa_TP3, Co2MnGa_ARPES}. A scaling analysis of the AHE as a function of temperature, combined with ARPES measurements of the Weyl lines and first-principles calculations, further suggested that Berry curvature concentrated by the Weyl lines drives the giant AHE in Co$_{2}$MnGa \cite{Co2MnGa_ARPES}. This represents a considerable success in achieving robust, high-temperature transport driven by a topological electronic structure. Fe$_3$Ga and Fe$_3$Al, which crystallize in the same space group as Co$_{2}$MnGa, were also shortly thereafter observed to host a giant ANE at room temperature \cite{Fe3Ga_Nakatsuji}. Through first-principles calculation, this giant ANE was attributed to Berry curvature generated by a rich Weyl line network extending throughout the bulk Brillouin zone. Apart from Co$_{2}$MnGa and the Fe$_3$Ga family, the van der Waals ferromagnet Fe$_3$GeTe$_2$ was found to exhibit a large AHE, which again was attributed to topological Weyl lines \cite{Fe3GeTe2_AHE_linenodes}. The kagome magnet Co$_{3}$Sn$_{2}$S$_{2}$ was also observed to exhibit a giant AHE, while first-principles calculations predicted that Co$_{3}$Sn$_{2}$S$_{2}$ hosts three topological Weyl lines in the absence of SOC \cite{Co3Sn2S21,Co3Sn2S22,Co3Sn2S23}. Under SOC, these Weyl lines gap out, leaving behind three pairs of Weyl points predicted at 60 meV above the Fermi level. Subsequent works explored the predicted Weyl points by ARPES or STM, relying on comparison with first-principles calculation to interpret certain bulk and surface states as signatures of Weyl points and topological Fermi arcs \cite{Co3Sn2S2_ARPES, Co3Sn2S2_STM2}. A large zero-field ANE was also reported \cite{Co3Sn2S2_ZeroFieldNernst}. At the same time, a large Berry curvature associated with the SOC-gapped Weyl lines was found at the Fermi level in calculations, while the predicted AHE exhibited its maximum value very close to the Fermi level and dropped steeply as one approached the Weyl point energy \cite{Co3Sn2S21}. As a result, some works suggested that the large AHE and ANE arise predominantly due to Berry curvature concentrated by the Weyl lines \cite{Co3Sn2S2_Shuang,Co3Sn2S2_ZeroFieldNernst,Co3Sn2S2_Belopolski}. More broadly, Co$_{3}$Sn$_{2}$S$_{2}$ has been a rich playground for a variety of phenomena bringing together the kagome lattice, magnetism and topology, such as negative flat-band magnetism \cite{Co3Sn2S2_STM}. The community has also given considerable attention to Fe$_{3}$Sn$_{2}$ and the Mn$_{3}$Sn family of topological magnets, which also exhibit a large AHE and ANE. However, for these materials, the Berry curvature is thought to be concentrated by massive Dirac fermions, in the case of Fe$_{3}$Sn$_{2}$ \cite{Fe3Sn2_STM,Fe3Sn2_ARPES}, and Weyl points, in the case of Mn$_{3}$Sn \cite{Mn3Ge1,Mn3Ge2, Mn3Ge3, Mn3Ge_DFT}. To determine the origin of the large AHE and ANE in topological magnets, spectroscopic measurements are a powerful tool. The future will no doubt see the discovery of many other examples of anomalous transport arising from topological electronic structures.

\bigskip

\textbf{VI. Outlook}

The materials discussed above represent only a small fraction of candidate topological conductors known to the community. As a helpful guide to the field, we attempt to list some of the existing predictions of Weyl materials, along with the current experimental status (Table I) \cite{WT_CA,TIT,TIT_ARPES,TIT_ARPES_2,TIT_Transport,WP2,WP2_transport,TlBiSe, Balents_HgTe,Vanderbilt,TeSe,TaS,TaS_Transport,CuTlSe2,BiX,Ag2Se,Korean_BiSb, HgCdTe, YbMnBi2,Heusler_Bernevig,Heusler_Hasan,CeSbTe,SrPtAs_WSC,SrPtAs_muSR,URu2Si2_WSC,UPt3_WSC,GdPtBi}. First-principles calculations have continued to play a crucial role in the search for new topological quantum materials in 2019. Some predictions have been experimentally confirmed, as the TaAs and RhSi families discussed here. Thanks to rapidly increasing computational power, material predictions have also greatly accelerated. For example, thousands of potential topological materials were predicted very recently \cite{catlog1,catlog2,catlog3}. At the same time, we note that identifying the most promising candidates involves avoiding the pitfalls of \textit{ab initio} calculations \cite{com2019}. Most of the predictions lack conclusive evidence, which indicates more  detailed calculation, as well as further experimental study, is needed in future. As highlighted by Zunger \cite{com2019}, it is important to check theoretically and experimentally that the material is stable in the desired crystal structure, possesses the required magnetic order, has sufficiently high crystalline quality, remains robust upon any required doping, is well-characterized in its electron-electron interactions, etc. Such considerations become even more important for topological phases relying on specific magnetic orders or other interacting phenomena, which can complicate \textit{ab initio} calculations. Going forward, the search for high-quality topological materials and exotic new topological states of matter will no doubt benefit more than ever from close collaboration across many powerful theoretical and experimental techniques.

One of the new frontiers in the exploration of topological materials is topological magnets with rich magnetic tunability. The ideal topological quantum materials host topological indices accompanied by a large Berry curvature field at the Fermi level. The interplay between the magnetic field and the giant Berry curvature field can induce many exotic quantum phenomena \cite{Mag_tune3, Mag_tune4}. Here we use kagome magnets as an example (Fig.~\ref{Fig8}\textbf{a}), whose electronic structure consists of a flat band and massless Dirac fermions (Fig.~\ref{Fig8}\textbf{b}). The first case is the anomalous magnetic response in a half-metallic kagome magnet Co$_{3}$Sn$_{2}$S$_{2}$  \cite{Co3Sn2S2_STM}. This material features a spin-polarized kagome flat band at the Fermi level, apparent as a sharp peak in the density of states (Fig.~\ref{Fig8}\textbf{c}). Under an external magnetic field, this flat band peak exhibits an anomalous energy shift, opposite to the applied field direction (Fig.~\ref{Fig8}\textbf{d}). This negative magnetic moment of the flat band can be understood from the underlying Berry curvature field.  As a second example,  the kagome magnets Fe$_{3}$Sn$_{2}$ exhibit a large and anisotropic magnetic field response  \cite{Fe3Sn2_STM}.
The system exhibits massive Dirac bands whose gap can be tuned by an external magnetic field (Fig.~\ref{Fig8}\textbf{e}) \cite{Fe3Sn2_ARPES}. The response has been found to be large, with an effective $g$-factor of over 100, two orders of magnitude larger than that of conventional electron spin. More unexpectedly, the electronic structure shows intricate nematicity, which can be systematically rotated by the vector magnetization (Fig.~\ref{Fig8}\textbf{f}). The exotic magnetic tunability of topological quantum materials offers a path to topologically protected spintronics applications in the areas of magnetic sensors and electromagnetic converters.

The observation of magnetic topological crystals has also opened new frontiers for quantized electromagnetic response. One example is the quantum anomalous Hall effect (AHE), a phenomenon in two-dimensional magnetic materials where the transverse conductivity is quantized to integer multiples of the conductance quantum, $e^2/h$. In existing setups, this effect persists to temperatures of only $\sim 1$ K and remains limited to doped thin films of compounds derived from the Bi$_2$Se$_3$ family of topological insulators \cite{HeWangXue_QAHE_ARCMP}. One approach to develop new quantum AHE materials considers thin films of magnetic Weyl semimetals, rather than topological insulators. The large anomalous Hall response observed in topological magnets \cite{AHE_Weyl,Co3Sn2S21, Co3Sn2S22, Fe3GeTe2_AHE_linenodes, Co2MnGa_ARPES, Co2MnGa_TP1, Co2MnGa_TP2, Co2MnGa_TP3,Fe3Sn2_ARPES,Mn3Ge1,Mn3Ge2, Mn3Ge3, Mn3Ge_DFT,Co3Sn2S2_STM,Co3Sn2S23,Co3Sn2S2_ARPES, Co3Sn2S2_STM2} also naturally motivates the search for a three-dimensional quantum AHE. This phase can be viewed as a stack of ordinary two-dimensional quantum AHE states or, alternatively, as a phase which arises when a magnetic Weyl semimetal is tuned in such a way that all the Weyl points have annihilated each other. A three-dimensional quantum AHE state may provide quantized conductance at high temperatures while circumventing the experimental challenges of working with a two-dimensional device. Lastly, it has recently been proposed that under certain conditions the linking number of the Weyl lines in an electronic structure may determine the quantized $\theta$ angle of the axion Lagrangian \cite{ChangYee_WeylLink}, suggesting that new topological mechanisms for quantized electromagnetic response still wait to be explored.

The field responses of the thin film of topological semimetals might open a new direction. Compared to bulk crystals, the thin film of topological materials, where surface states may be dominating, can have different electronic, thermal, magnetic, and optical properties. For example, subject to external magnetic fields, Fermi arcs on opposite surfaces of the sample play a key role in the quantum Hall effect of the Cd$_{3}$As$_{2}$ thin films \cite{SdH_exp2}. Under circularly polarized laser, the injection currents from Fermi arcs may also be crucial in the thin film of the topological chiral crystal RhSi \cite{arc_photo}. 

Novel superconducting and correlated topological phases also lie on the horizon. In RhSi, the large separation of the high-fold chiral fermions in momentum space may offer the possibility to engineer new kinds of unconventional superconductivity via doping or proximity effect \cite{Weyl-SC-1, Weyl-SC-2, Weyl-SC-5, Weyl-SC-7, sup_chiral1}.  Magnetic topological materials including kagome flat bands may also allow to explore the interacting topological phases, including the fractional topological states. The further theoretical and experimental foundation needed to realize such exotic states of matter will no doubt emerge as an important future direction for the field.

\clearpage
\textbf{Box 1:  Topological invariants and Weyl fermions}

Topological invariants are integer indices that define the topology and determine the exotic properties of topological quantum materials \cite{Graphene, Hasan2010, Qi2011, TI_book_2014_2, Bansil_rev}.  These quantized invariants are rooted in the Berry phase  $\gamma_{m}$ \cite{Mag_tune3} which is acquired by a path integral in momentum space from the Bloch wave functions $| m (\boldsymbol{k}) \rangle$ over evolution of the Hamiltonian $\mathcal{H}(\boldsymbol{k})$:

\begin{eqnarray}
\gamma_{m}= \int_{c} \mathcal{A}_{m}(\boldsymbol{k}) \cdot d\boldsymbol{k}
\end{eqnarray}

where $\mathcal{A}_{m}(\boldsymbol{k})$ is a vector-valued function called the Berry connection.
\begin{eqnarray}
\mathcal{A}_{m}(\boldsymbol{k})=i \langle m (\boldsymbol{k}) | \frac{\partial }{\partial \boldsymbol{k}} | m(\boldsymbol{k}) \rangle
\end{eqnarray}

The integrated Berry connection along any path that circling a topological nodal line is a quantized integer  $\gamma=\pi$ (Fig.~\ref{box1}\textbf{a}). The quantized Berry phase of a nodal line is the same as the winding number for gapless points in two-dimensional systems, such as the massless Dirac fermions in graphene \cite{Graphene}.

The topological invariant of three-dimensional Weyl fermions are associated with the Berry curvature field:

\begin{eqnarray}
\boldsymbol{\Omega}_{m} (\boldsymbol{k})=\nabla_{\boldsymbol{k}} \times \mathcal{A}_{m}(\boldsymbol{k})
\end{eqnarray}
 Weyl fermions behaves as source or sink of Berry curvature field ( Fig.~\ref{box1}\textbf{b}). The integral of the Berry curvature field of any two-dimensional manifold enclosing a Weyl point is an integer, the Chern number or chiral charge of the Weyl point:
\begin{eqnarray}
C_{m}= \frac{1}{2\pi} \int_{S} \boldsymbol{\Omega}_{m} (\boldsymbol{k}) d\boldsymbol{S}
\end{eqnarray}
The Berry curvature field has the same transformation properties as the magnetic field under inversion and time-reversal symmetry. In this sense, the Berry curvature field can be viewed as the magnetic field in momentum space. Therefore Weyl fermions can be interpreted as magnetic monopoles in momentum space. Due to the quantized chiral charge of each Weyl node, meeting two oppositely-charged Weyl fermions in momentum space is the only way to annihilate Weyl fermions in a periodic system.

\begin{center}
\begin{table*}
\begin{tiny}
\centering
\begin{tabular}{P{2.1cm}P{2.7cm}P{3cm}P{3.5cm}P{3.7cm}}
\hline
Materials & Theoretical Predictions & Spectroscopy Measurements & Magnetic and Optical Responses in Experiments & Remarks \\
\hline
\hline
\textbf{TaAs family} & $\mathcal{I}$-breaking;  24 Weyl fermions \cite{Huang2015, Weng2015} & ARPES and STM \cite{TaAs_Hasan, TaAs_Ding, NbAs_Hasan, TaAs_Ding_2, TaAs_Chen, TaP_Hasan, TaP_Shi, NbP_Hasan, TaP_Chen,  NbP_STM1, TaAs_STM1, TaAs_STM2,spin1,spin2} & Negative longitudinal magneto-resistance \cite{Chiral_anomaly_ChenGF, Chiral_anomaly_Jia, NbAs_Transport_1, NbAs_Transport_2, NbP_MAG, TaP_Transport_1,TaP_Transport_2, TaP_Chiral_anomaly_1}; Nonlinear optical responses \cite{TaAs_OP2, TaAs_optics,TaAs_exp2} & Affirmative in experiments; First Weyl semimetal \\
\\
\textbf{RhSi family} & $\mathcal{I}$-breaking; 2 high-fold chiral fermions  \cite{KramersWeyl, RhSi, CoSi, FeSi} & ARPES  \cite{RhSi_exp, CoSi_exp, AlPt_exp, CoSi_exp2,FeSi_exp} & Quantized circular photogalvanic effects \cite{CPGE_exp_Q} & Affirmative in experiments as nearly ideal Weyl semimetals\\
\\
 LaAlGe & $\mathcal{I}$-breaking; 40 Weyl fermions \cite{LAG} & ARPES \cite{LAG} & Not reported & Type-II Weyl cones observed; Fermi arcs data is needed\\
 \\
 W$_{1-x}$Mo$_x$Te$_2$ & $\mathcal{I}$-breaking; 8 Weyl fermions \cite{WT-Weyl,WMoTe-Weyl, MT-Weyl}  &  ARPES and STM \cite{WT-ARPES-1, WT-ARPES-2, WT-ARPES-3, WT-ARPES-4, WT-ARPES-5, WT-ARPES-8, WT-STM-1,WT-STM-2} & Negative longitudinal magneto-resistance \cite{WT_CA} & Weyl fermions above the Fermi level in predictions; Clear type-II bulk cone data is needed \\

 \hline

 TaIrTe$_4$  & $\mathcal{I}$-breaking; 4 Weyl fermions \cite{TIT} &  ARPES \cite{TIT_ARPES, TIT_ARPES_2} &  Preliminary transport \cite{TIT_Transport}; Nonlinear optical responses \cite{MoTephoto_exp} & Weyl fermions above the Fermi level in predictions; Clear type-II bulk cone data is needed\\
 \\
 WP$_2$, MoP$_2$ & $\mathcal{I}$-breaking; 8 Weyl fermions \cite{WP2} & ARPES \cite{WP2_ARPES} & Preliminary transport \cite{WP2_transport} & Direct experimental evidence of Weyl fermions is needed\\
 \\
 PrAlGe, CeAlGe  & $\mathcal{T}$-breaking \cite{RAlX} & ARPES \cite{CeAlGe_ARPES} & Singular angular magnetoresistance \cite{CeAlGe_transport} &  \\
\\
Mn$_3$Ge, Mn$_3$Sn & Noncollinear antiferromagnic Weyl \cite{Mn3Ge_DFT}  &  ARPES \cite{Mn3Ge3} &  Strong anomalous Hall effect \cite{Mn3Ge1, Mn3Ge2} & Direct observation of Weyl fermions and Fermi arcs needed \\
\\
Co$_{3}$Sn$_{2}$S$_{2}$ & Noncollinear antiferromagnic Weyl \cite{Co3Sn2S21, Co3Sn2S22,Co3Sn2S23,Co3Sn2S2_ARPES,Co3Sn2S2_STM2} & ARPES and STM \cite{Co3Sn2S21, Co3Sn2S22,Co3Sn2S23}  & Strong anomalous Hall effect \cite{Co3Sn2S21, Co3Sn2S22}  & Kagome flatbabds detected \cite{Co3Sn2S2_STM}; Weyl fermions above the Fermi level in predictions\\
\\
TlBi(S$_{0.5}$Se$_{0.5}$)$_{2}$  & $\mathcal{I}$-breaking \cite{TlBiSe}  & Not reported & Not reported &  \\
\\
HgTe/CdTe superlattice & $\mathcal{I}$-breaking; 8 Weyl fermions \cite{Balents_HgTe} & Not reported & Not reported &  \\
\\
LaBi$_{1-x}$Sb$_{x}$Te$_{3}$ family  & $\mathcal{I}$-breaking; 12 Weyl fermions \cite{Vanderbilt}  & Not reported & Not reported &  \\
\\
BiTeI & $\mathcal{I}$-breaking;12 Weyl fermions \cite{Vanderbilt}  & Not reported & Not reported &  \\
\\
Te \& Se in the trigonal phase & $\mathcal{I}$-breaking \cite{TeSe} & Not reported & Not reported &  High pressure required \\
\\
SrSi$_2$ & $\mathcal{I}$-breaking; Double Weyl   \cite{SrSi2} & Not reported  & Not reported & \\
\\
Ta$_3$S$_2$  & $\mathcal{I}$-breaking; 40 Weyl fermions \cite{TaS} & Not reported & Preliminary transport \cite{TaS_Transport} & \\

\hline
\\
\end{tabular}

\end{tiny}
\end{table*}
\end{center}

\addtocounter{table}{-1}
  \begin{center}
\begin{table*}
\begin{tiny}
\centering
\begin{tabular}{P{2.1cm}P{2.7cm}P{3cm}P{3.5cm}P{3.7cm}}
\hline
Materials & Theoretical Predictions & Spectroscopy Measurements & Magnetic and Optical Responses in Experiments  & Remarks \\
\hline
\hline
CuTlSe$_2$ family & $\mathcal{I}$-breaking; 8 Weyl fermions \cite{CuTlSe2}  & Not reported & Not reported &  \\
\\
$\beta$-Bi$_{4}$Br$_{4}$ family  & $\mathcal{I}$-breaking; 4 Weyl fermions \cite{BiX} & Not reported & Not reported &  High pressure required \\

\\Ag$_{2}$Se & $\mathcal{I}$-breaking; Kramers-Weyl \cite{KramersWeyl} & Not reported & Negative longitudinal magneto-resistance \cite{Ag2Se} & \\
\\
R$_2$Ir$_2$O$_7$& $\mathcal{T}$-breaking; 24 Weyl fermions \cite{Wan2011}  & Not reported & Not reported& Magnetic structure unsettled \cite{Iridate_Mag_1, Iridate_Mag_2} \\
\\
TI/FM insulator superlattice& $\mathcal{T}$-breaking; 2 Weyl fermions\cite{Burkov2011}  & Not reported & Not reported &  \\
\\
HgCr$_2$Se$_4$ & $\mathcal{T}$-breaking; double Weyl \cite{HgCrSe} & Not reported & Not reported &  \\
\\
Bi$_{0.97}$Sb$_{0.03}$  & $\mathcal{T}$-breaking \cite{Korean_BiSb} & Not reported& Negative longitudinal magneto-resistance\cite{Korean_BiSb} & External magnetic field required  \\
\\
Hg$_{1-x-y}$Cd$_{x}$Mn$_{y}$Te & $\mathcal{T}$-breaking; 2 Weyl fermions \cite{HgCdTe} & Not reported & Not reported & External magnetic field required  \\
\\
YbMnBi$_2$ & $\mathcal{T}$-breaking \cite{YbMnBi2}  & ARPES \cite{YbMnBi2} & Not reported & Magnetic order needs further confirmation \\
 \\
 CeSbTe & $\mathcal{T}$-breaking \cite{CeSbTe} & Not reported & Preliminary transport \cite{CeSbTe} & Direct observation of Weyl fermions and Fermi arcs needed \\
\\
GdPtBi & $\mathcal{T}$-breaking \cite{GdPtBi}  & Not reported  & Negative longitudinal magneto-resistance\cite{GdPtBi} & External magnetic field required \\
 \\
Magnetic Heusler materials XCo$_{2}$Z (X=IVB or VB; Z=IVA or IIIA) & $\mathcal{T}$-breaking  \cite{Heusler_Hasan, Heusler_Bernevig}  & Not reported & Not reported &  \\
\\
SrPtAs & $\mathcal{T}$-breaking; Superconducting \cite{SrPtAs_WSC} & Not reported &  Not reported& $\mu$SR suggested a $\mathcal{T}$-breaking SC pairing; $T_c=2.4$ K \cite{SrPtAs_muSR} \\
 \\
URu$_2$Si$_2$ & $\mathcal{T}$-breaking; 2 Weyl fermions; Superconducting \cite{URu2Si2_WSC} & Not reported & Not reported &  \\
\\
B-phase of UPt$_3$ & $\mathcal{T}$-breaking; Superconducting \cite{UPt3_WSC} & Not reported & Not reported &  \\
\\
Li$_{2}$Pd$_{3}$B & $\mathcal{I}$-breaking; Kramers-Weyl Superconducting \cite{KramersWeyl} & Not reported & Not reported &  \\
\hline
\\
\end{tabular}
\end{tiny}
\caption{\textbf{Theoretical prediction and experimental investigation of Weyl material candidates.} A list of predicted inversion-breaking ($\mathcal{I}$-breaking) and time-reversal-breaking ($\mathcal{T}$-breaking) Weyl semimetal candidates. Most of them lack conclusive experimental evidence. }
\label{Table I}
\end{table*}
\end{center}

\clearpage

\begin{figure*}
\centering
\includegraphics[width=16cm]{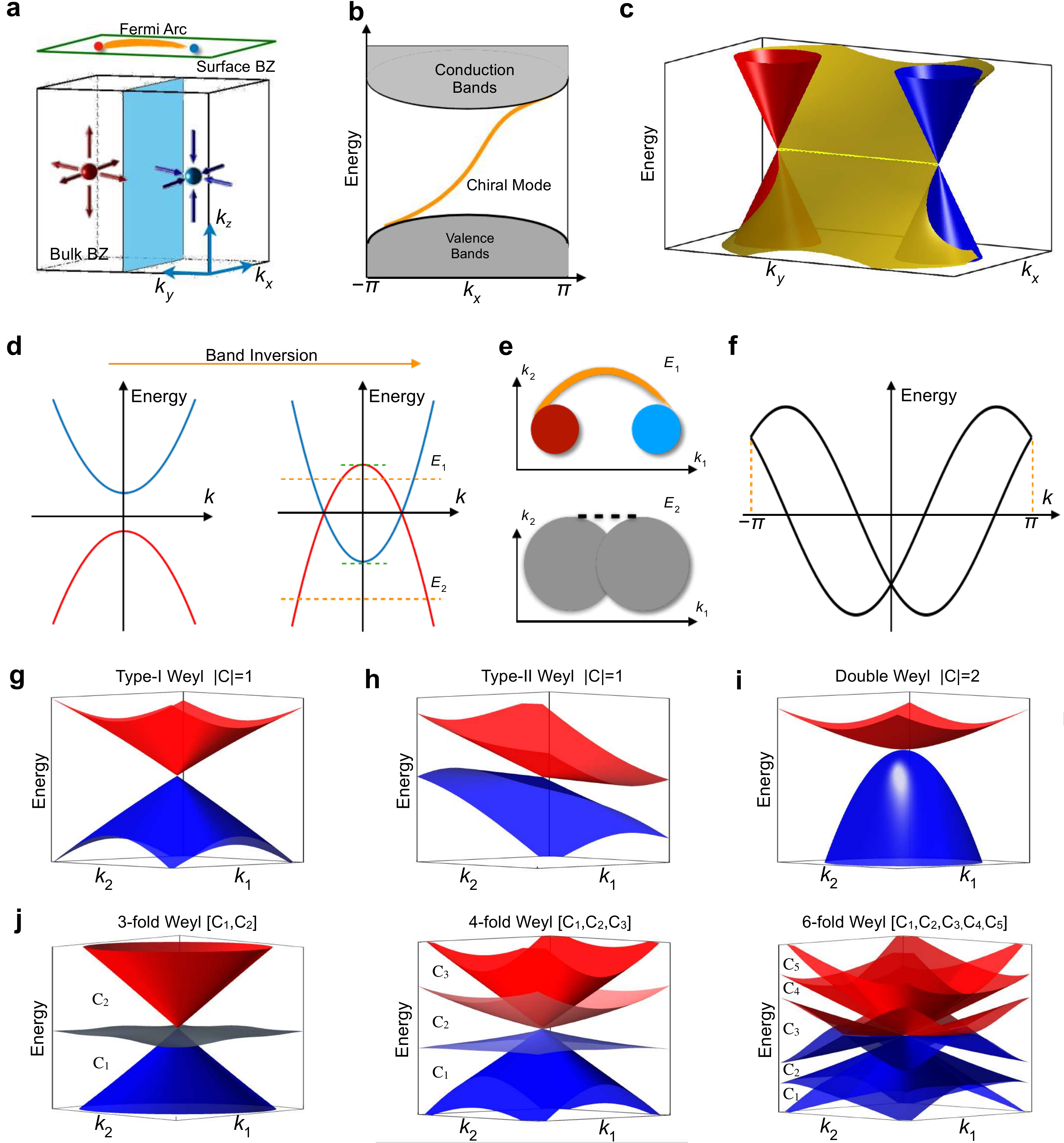}
\caption{\label{Fig1}\textbf{Topology and classification of Weyl fermions} \textbf{a,} Schematic of two Weyl fermions in bulk  and the Fermi arc on the surface. The red and blue balls are Weyl fermions of opposite chiral charges. The arrows show the flow of the Berry curvature field.   \textbf{b,} The blue plane in (a) in between two Weyl fermions  is a two-dimensional subsystem with a quantized Chern number, inducing a chiral edge mode connecting the conduction band and valence band. \textbf{c,} Schematic of the Fermi arc surface state (the yellow surface)  between two oppositely-charged Weyl fermions (red and blue cones) in energy-momentum space.}
\label{Fig1}
\end{figure*}

\addtocounter{figure}{-1}
\begin{figure*}[t!]
\caption{ \textbf{d,}  Weyl fermions can be generated via band inversion in crystals breaking of time-reversal or inversion symmetry. The two-fold band crossings between the red and blue bands are two Weyl fermions of opposite chiral charges. The topologically nontrivial energy window is within the top of the red band and the bottom of the blue band (the green dashed line) \textbf{e,} Bulk pockets and surface states at different energies in a Weyl semimetal. Top: Within the topologically nontrivial energy window, the bulk pockets of opposite chiral charges  are isolated. Topological Fermi arcs (yellow line) connects projected bulk pockets on the surface. Bottom: Out of the topologically nontrivial energy window, two Weyl fermions merge into one trivial (grey) pocket. No topological Fermi arc is protected.  \textbf{f,}  Without band inversion, Kramers-Weyl fermions in non-magnetic chiral crystals are created by spin-orbit coupling (SOC).  \textbf{g-i,} Classification of different two-fold Weyl fermions by their $E-k$ dispersions and the topological chiral charges. \textbf{j,} Energy dispersions of high-fold chiral fermions. The topological invariants of a $n$-fold chiral fermions can be defined by the Chern numbers of the $n$-1 band gaps. }
\label{Fig1}
\end{figure*}

\clearpage

\begin{figure*}
\centering
\includegraphics[width=16cm]{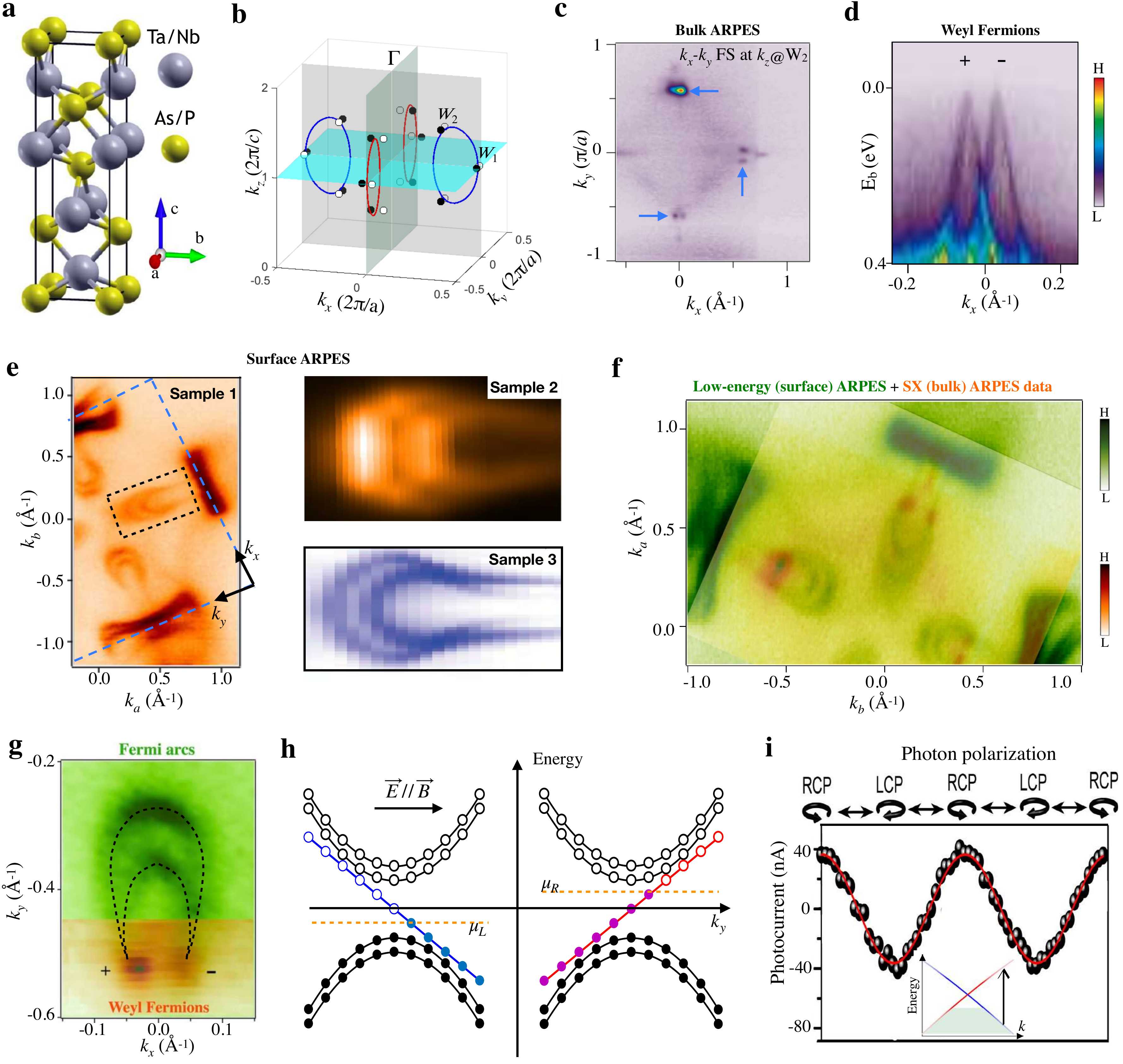}
\caption{\label{Fig2}\textbf{Weyl semimetal states in the noncentrosymmetric crystals the TaAs family} \textbf{a,} Inversion symmetry breaking crystal structure of the TaAs family. The lattice constants are a = 3.437 $\rm \AA$ and c = 11.656 $\rm \AA$. \textbf{b,} Nodal degeneracies of TaAs in momentum space.  In the absence of SOC, there are four nodal rings on the mirror planes $k_{x}=0$ and $k_{x}=0$. After turning on SOC, two groups of Weyl fermions created: $W_{1}$  and $W_{2}$.  The white and black balls denote Weyl fermions of opposite chiral charges.   \textbf{c,} SX-ARPES-measured  $k_{x}-k_{y}$ Fermi surface map of the $W_{2}$ Weyl node.    \textbf{d,} SX-ARPES-measured $E-k_{x}$ dispersion cutting through two $W_{2}$  Weyl fermions. Two linearly dispersive Weyl cones are observed. The color bar limits correspond to high (H) and low (L) electron density.}
\label{Fig2}
\end{figure*}

\addtocounter{figure}{-1}
\begin{figure*}[t!]
\caption{ \textbf{e,} ARPES-measured (001) surface electronic structures of TaAs from different groups. The crescent-shaped features are the Fermi arcs.  \textbf{f,} UV-ARPES-measured Fermi surface map (surface states) with the SX-ARPES map (bulk states) overlaid at the same scale shows that the locations of the projected bulk Weyl fermions correspond to the terminations of the surface Fermi arcs.  \textbf{g,} High-resolution ARPES maps of the Fermi arcs and the Weyl fermion nodes. The black dashed lines are the traces of Fermi arcs.   \textbf{h,} The Chiral anomaly effect of Weyl semimetals. Under parallel electric and magnetic fields, electrons are pumped from one Weyl fermion to the oppositely-charged one.   \textbf{i,} Polarization-dependent photocurrents at T = 10 K measured  in TaAs. The asymmetric excitation of a Weyl fermion under circularly polarized laser is illustrated in the inset.  Adapted from Ref. \cite{TaAs_Hasan, Huang2015, TaAs_optics, TaAs_Ding, Weng2015, TaAs_Chen}. Detailed first-principles predictions can be found in Ref. \cite{Huang2015, Weng2015}. Comprehensive ARPES-results of the TaAs family can be found in Ref. \cite{TaAs_Hasan, NbAs_Hasan, TaAs_Ding,TaAs_Ding_2, TaAs_Chen, TaP_Hasan, TaP_Shi, NbP_Hasan, TaP_Chen}.}
\label{Fig2}
\end{figure*}

\clearpage

\begin{figure*}
\centering
\includegraphics[width=16cm]{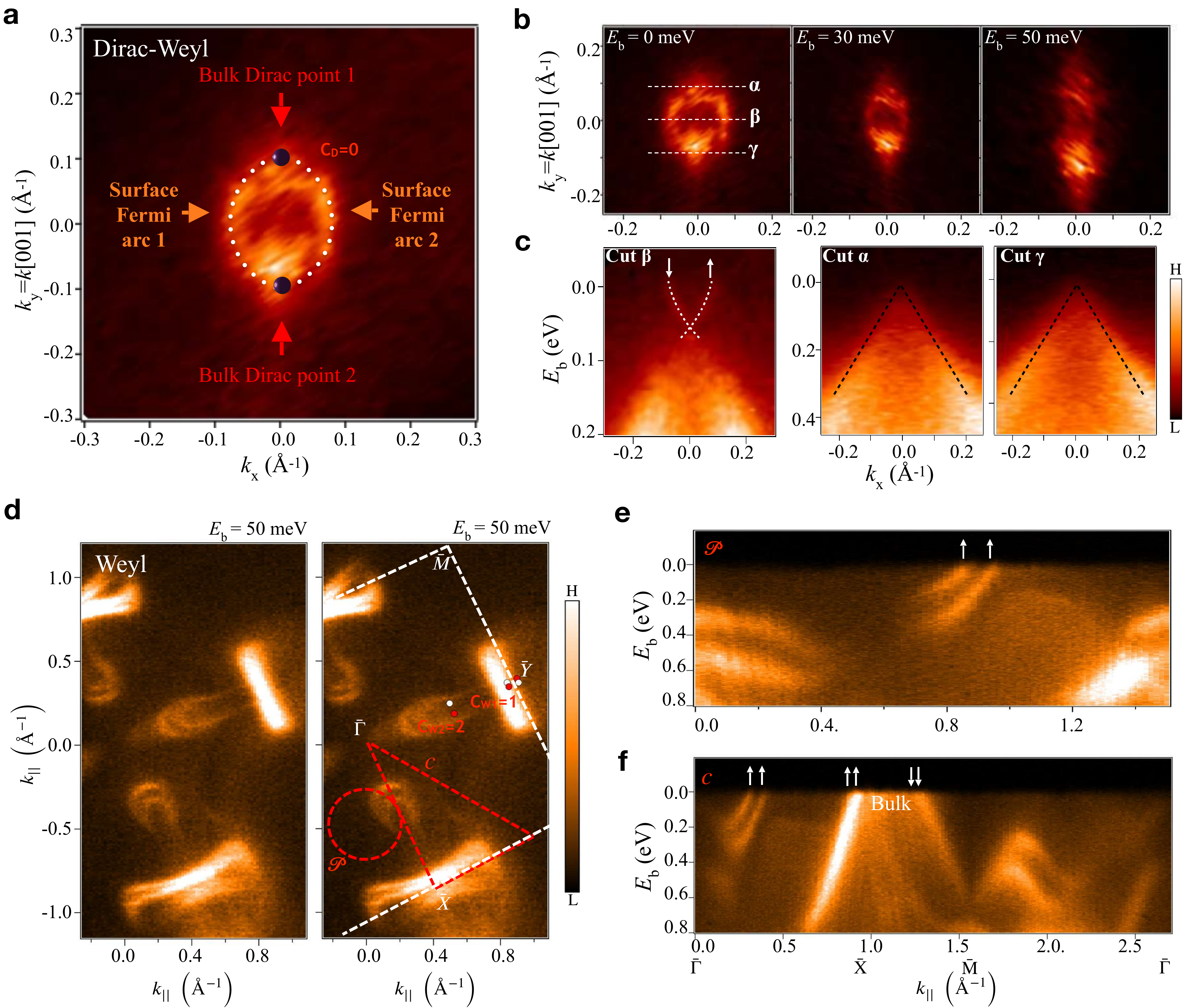}
\caption{\label{Fig3}\textbf{Chern numbers and Fermi arc surface states} \textbf{a,} ARPES-measured (100) surface states of the Dirac semimetal Na$_{3}$Bi. Two Fermi arcs (white dotted lines) connecting to bulk Dirac points (black balls, $C=0$)  form a closed loop.  \textbf{b,} Surface states of  Na$_{3}$Bi at different binding energies. \textbf{c,} ARPES-measured energy dispersions along the paths $\alpha$, $\beta$, and $\gamma$ illustrated in panel (b). The white dotted lines indicate the surface Fermi arcs with their chiralities marked by arrows. The black dashed lines indicate the bulk cones.  \textbf{d,} (001) surface states of TaAs at the energy of 50 meV below the Fermi level. The projected Weyl fermions on the surface are indicated by the red and white balls in the right panel, and the boundary of the BZ is indicated by the white dashed line. \textbf{e,} The energy dispersions along the red loop $p$. Two chiral co-propagating states are observed, consistent with net charge $C=2$ in the loop. \textbf{f,} The energy dispersions along the red triangle path $c$. Bulk states pass through the Fermi level along $\bar{X}$-$\bar{M}$ making it difficult to determine the the chiral surface states and related Chern numbers.  Adapted from Ref. \cite{Hasan_Na3Bi, NbP_Hasan}. }
\label{Fig3*}
\end{figure*}

\begin{figure*}
\centering
\includegraphics[width=16cm]{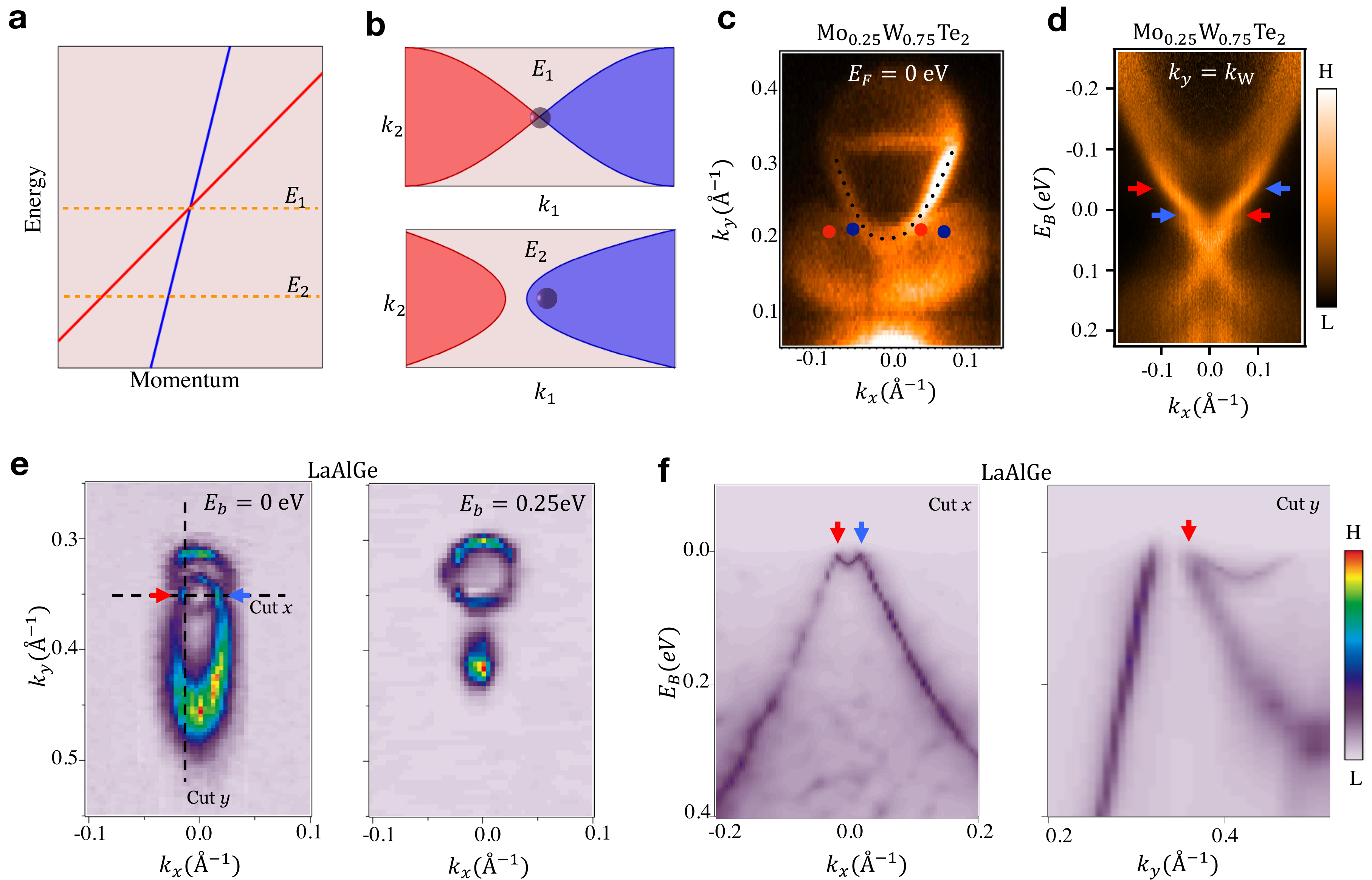}
\caption{\label{Fig3}\textbf{Type-II Weyl semimetals: the LaAlGe family and the Mo$_{x}$W$_{1-x}$Te$_{2}$ family} \textbf{a,} Schematic of a type-II Weyl fermion. \textbf{b,} Constant energy contours of a type-II Weyl fermion at different energies. The red (blue) region is the electron (hole) pocket. The black ball indicates the location of the Weyl node.  \textbf{c,} ARPES-measured Fermi surface of Mo$_{0.25}$W$_{0.75}$Te$_{2}$. The black dots are the trace of the topologically trivial surface arc. The red and blue balls are the predicted Weyl fermions on the surface. \textbf{d,} The energy dispersion cutting through the expected location of the Weyl points, showing a kink-like feature above the Fermi level that may be consistent with a Fermi arc. \textbf{e,} Constant energy contours of the type-II Weyl fermions in LaAlGe.  \textbf{f,} Bulk energy dispersions of the type-II Weyl cone in LaAlGe.  Adapted from Ref. \cite{WT-ARPES-4, LAG}. Detailed ARPES data of WTe$_{2}$ and MoTe$_{2}$ can be found in Ref. \cite{WT-ARPES-1, WT-ARPES-2, WT-ARPES-3, WT-ARPES-4, WT-ARPES-5,  WT-ARPES-8,WT-ARPES-9}. More ARPES data of LaAlGe can be found in Ref. \cite{LAG}.}
\label{Fig4}
\end{figure*}

\clearpage

\begin{figure*}
\includegraphics [width=16cm]{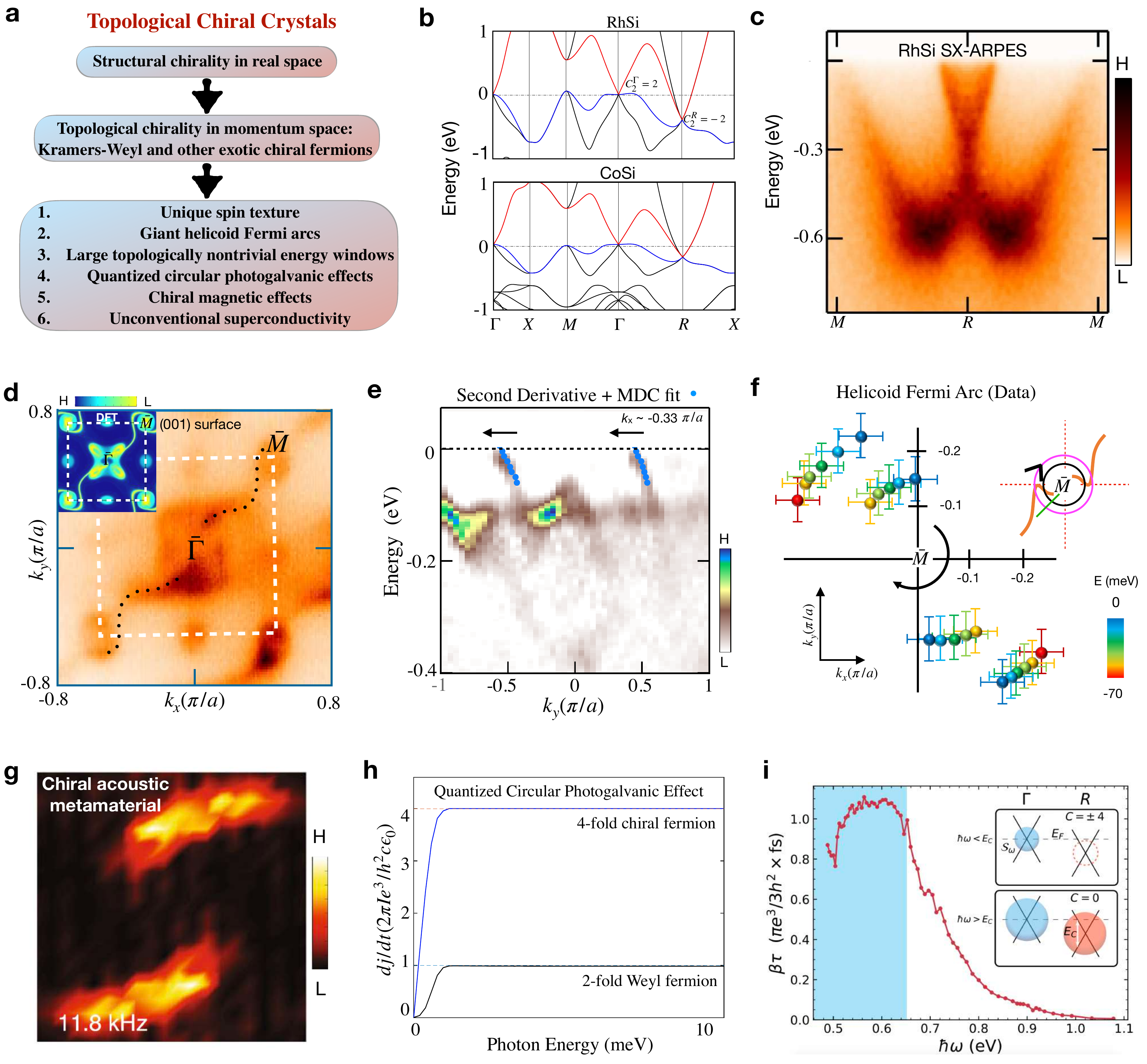}
\caption{\label{Fig5}\textbf{Topological chiral crystals in the RhSi family} \textbf{a,} Theoretically predicted exotic properties of topological chiral crystals. \textbf{b,} Electronic structures of CoSi and RhSi in the chiral space group 198 in the absence of SOC. The Chern numbers of the three-fold chiral fermions  are $C^{\Gamma}$=[2,2].  \textbf{c,} SX-ARPES-measured bulk cone of RhSi at the $R$ point. \textbf{d,} ARPES-measured Fermi surfaces with BZ boundary marked (white dashed line). Giant Fermi arc surface states extending diagonally across the BZ are observed in ARPES, consisting with DFT calculations on left-top. The black dotted lines are the traces of Fermi arcs. \textbf{e,} The ARPES-measured energy dispersion demonstrates the chirality of the Fermi arcs in the RhSi family. \textbf{f,}  The rotation of Fermi arcs along loops around $\bar{M}$ point with the changing of energy demonstrates the helicoid structure of Fermi arcs.  \textbf{g,}  Measured surface iso-frequency contours of chiral acoustic metamaterial.}
\label{Fig5}
\end{figure*}

\addtocounter{figure}{-1}
\begin{figure*}[t!]
\caption{  \textbf{h,} Numerical simulations of quantized circular photocurrents from a two-fold Weyl fermion and a four-fold chiral fermions. $I$ is the applied intensity of laser. \textbf{i, } CPGE amplitude as a function photon energy. The plateau at low energy were observed, which are consistent with theoretical predictions. Under higher photon energy (0.66 eV above), the photocurrent decays rapidly to 0. The red arrow on the inset indicated blocked optical transition under lower photon energy. Adapted from Ref. \cite{KramersWeyl, RhSi, RhSi_exp,CoSi_exp, AlPt_exp,acoustic1,CoSi,CoSi_exp2,CPGE_exp_Q,Moore2}. The universal theory of non-magnetic chiral crystals can be found in Ref. \cite{KramersWeyl}. First-principles predictions of ideal chiral crystals in the space group 198  can been found in Ref. \cite{RhSi, CoSi}. The spectroscopy evidence of topological chiral crystals is shown in Ref. \cite{RhSi_exp, CoSi_exp, AlPt_exp, CoSi_exp2, acoustic1}. The detailed studies of acoustic chiral crystals can be found in \cite{acoustic1,acoustic2,FeSi,FeSi_exp}. Theoretical study of quantized CPGE from two-fold Weyl fermions can be found in Ref. \cite{Moore2}. The predictions of quantized from high-fold chiral fermions can be found in Ref. \cite{RhSi}. The experimental realization of  quantized CPGE can be found in Ref. \cite{CPGE_exp_Q}.}
\label{Fig5}
\end{figure*}

\clearpage

\begin{figure*}
\includegraphics [width=16cm]{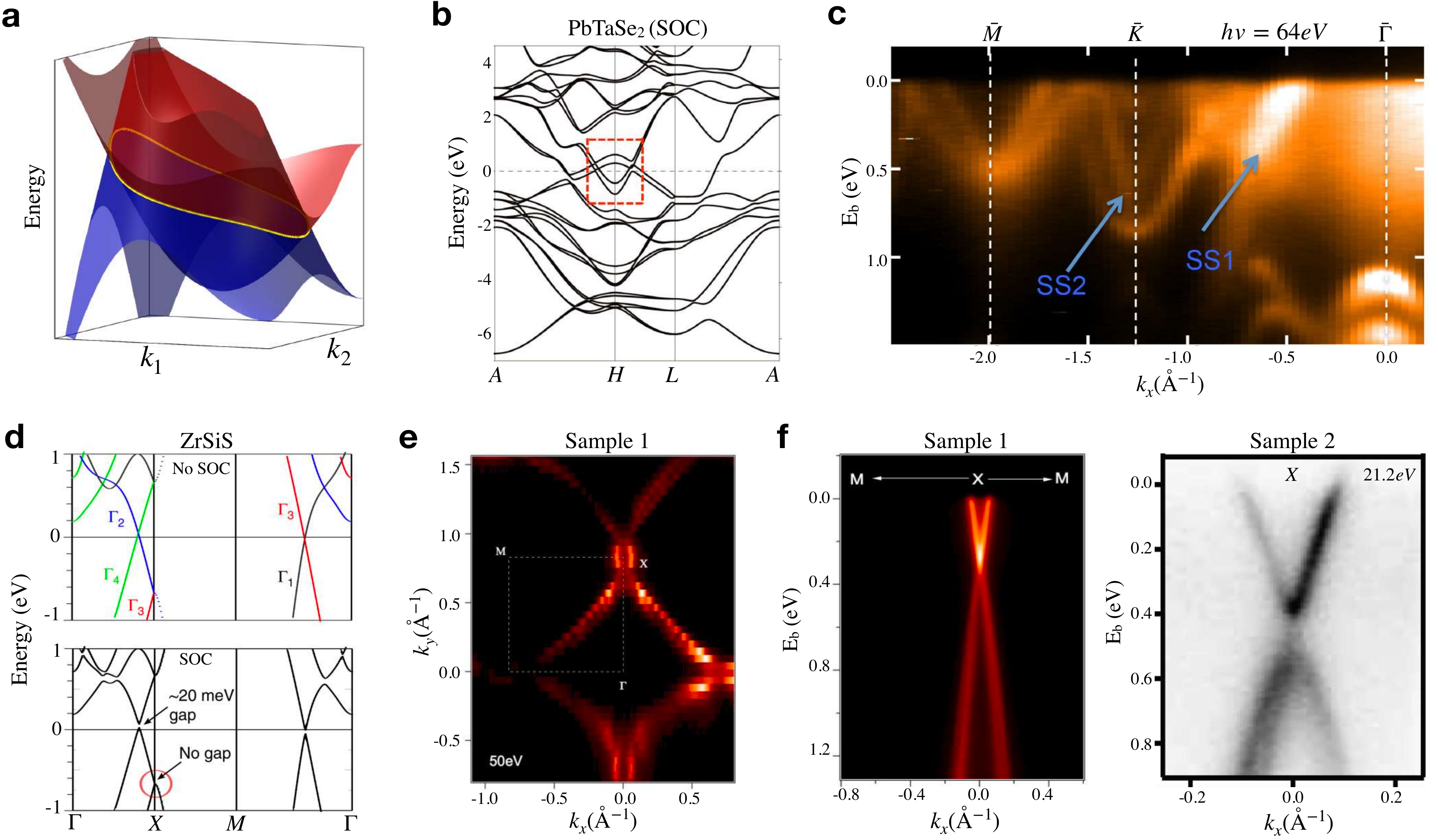}
\caption{\label{Fig6}\textbf{Topological nodal lines in PbTaSe$_{2}$ and the ZrSiX family.} \textbf{a,} Energy-dispersions of nodal line semimetals with one-dimensional nodal degeneracies (the yellow loop).   \textbf{b,}  Calculated electronic structures of PbTaSe$_{2}$ in the presence of SOC. \textbf{c,} ARPES-observed potential drumhead surface states (SS) in PbTaSe$_{2}$ are indicated by the arrows.   \textbf{d,}  Band structures of ZrSiS in the absence of SOC (top panel) and in the presence of SOC (bottom panel).  \textbf{e,} ARPES-measured constant energy contours of ZrSiS.   \textbf{f,}  Energy dispersions of ZrSiS in different samples show clear bulk cones of the nodal line.  Adapted from Ref. \cite{PTS, ZrSiS_NC, ZrSiS}. Further experimental study on ZrSiSe and ZrSiTe can be found in  \cite{ZrSiSe_ZrSiTe,ZrSiSe_ZrSiTe_Neupane,ZrSiSe_Hao}. }

\end{figure*}

\begin{figure*}
\includegraphics [width=13.5cm]{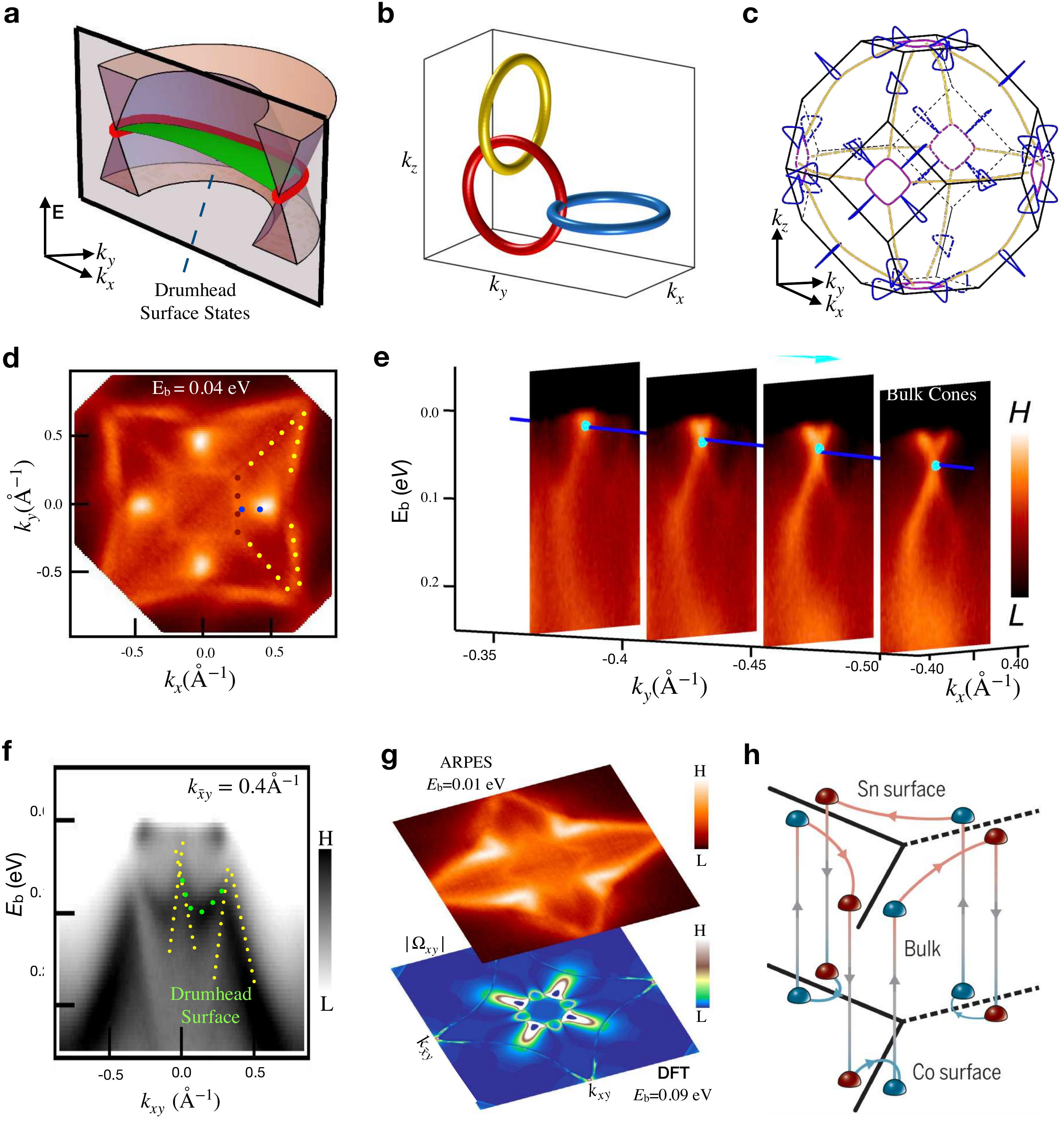}
\caption{\label{Fig7}\textbf{Magnetic Weyl lines and drumhead surface states in Co$_{2}$MnGa} \textbf{a,}  Schematic of a nodal line and its drumhead surface state. \textbf{b,}  An example of a nodal network built from multiple vertical Weyl rings. \textbf{c,} DFT predictions of the Weyl-line network in the momentum space of Co$_{2}$MnGa. \textbf{d,} ARPES-measured Fermi surfaces at $E_{b}$=0.04 eV. The colored dots are related to the different Weyl lines in panel c.  \textbf{e,}  Series of ARPES-measured Weyl-line cones along different k-paths.  \textbf{f,}  ARPES-measured energy dispersion along $k_{\bar{x}y}$ = 0.4 \AA$^{-1}$. The yellow dots illustrate the Weyl cones of the Weyl lines. The additional states connecting the Weyl line nodes are the drumhead surface states (green dots).  \textbf{g,} Top: ARPES-measured constant-energy contour at $E_{b}=0.01~eV$  on the (001) surface. Bottom: The calculated Berry curvature field intensity. \textbf{h,} Schematic of  three pairs of Weyl fermions and their Fermi arcs in Co$_{3}$Sn$_{2}$S$_{2}$. Adapted from Ref. \cite{Co2MnGa_theory,Co2MnGa_ARPES,Comment_mag}. Detailed spectroscopic results on Co$_{3}$Sn$_{2}$S$_{2}$ can be found in Ref. \cite{Co3Sn2S21, Co3Sn2S22,Co3Sn2S23,Co3Sn2S2_STM,Co3Sn2S2_ARPES,Co3Sn2S2_STM2}.}

\end{figure*}

\begin{figure*}
\includegraphics [width=17cm]{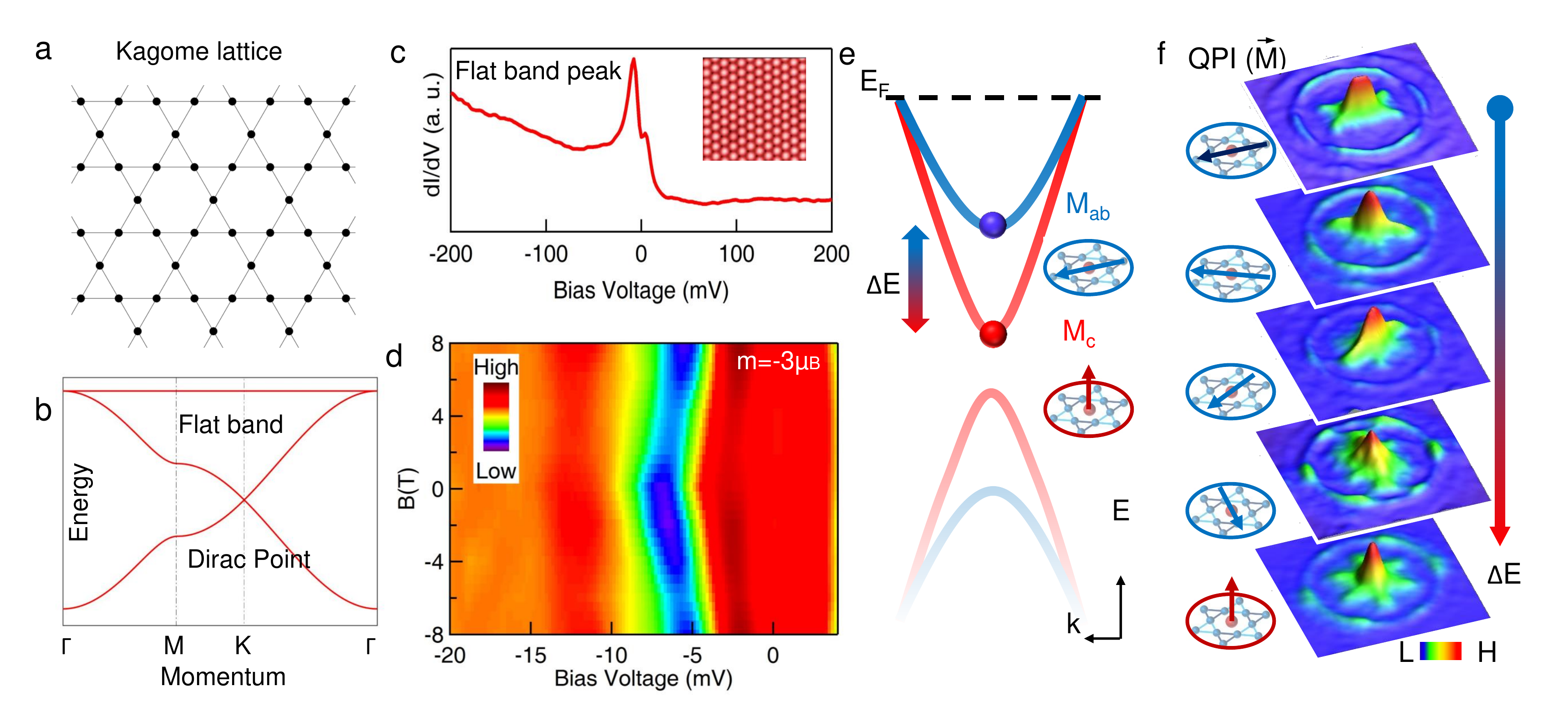}
\caption{\label{Fig8}\textbf{Magnetic tunability of topological Kagome magnets}   \textbf{a,} Illustration of a kagome lattice. \textbf{b,} Band structure of a kagome lattice with a flat band and a Dirac point. \textbf{c,} Tunnelling spectrum on the S surface (inset) of Co$_{3}$Sn$_{2}$S$_{2}$ showing the pronounced kagome flat band peak. \textbf{d,} C-axis field response of the flat band peak. The peak exhibits a shift in the energy of equal magnitude with the field applied in either c-axis direction. The derived magnetic moment value is around $m = -3 ~\mu B$. \textbf{e,} Schematic of the magnetization-controlled Dirac gap from the Kagome lattice. \textbf{f,} QPI patterns  of Fe$_{3}$Sn$_{2}$ as a function of the magnetization direction, which is indicated in the insets with respect to the kagome lattice. The uppermost QPI pattern shows the spontaneous nematicity along the $a$ axis. Magnetization along other directions can alter, and thus control, the electronic symmetry. Adapted from Ref. \cite{Fe3Sn2_STM, Co3Sn2S2_STM}. The detailed ARPES observation of massive Dirac fermions in Fe$_{3}$Sn$_{2}$ can been found in Ref. \cite{Fe3Sn2_ARPES}.}
\label{Fig8}
\end{figure*}

\makeatletter
\renewcommand{\thefigure}{B1}
\makeatother

\begin{figure*}
\centering
\includegraphics[width=16cm]{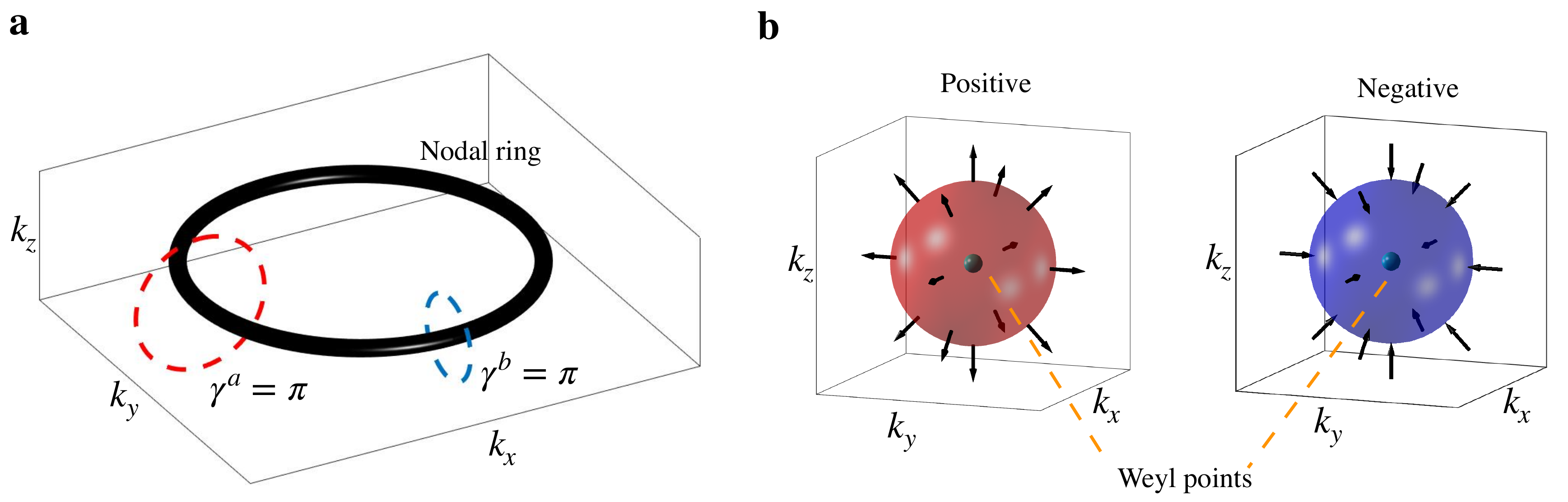}
\caption{\label{box1}\textbf{Topological invariants of massless fermions} \textbf{a,} The winding number of a topological nodal line is the quantized value $\pi$. The big black circle is a topological nodal line in momentum space, and the red and blue dashed lines are two different paths circling the rings.  \textbf{b,}  A Weyl fermion is a source or sink of the Berry curvature field (black arrows). The integral of the Berry curvature field of any two-dimensional manifold (such as the red or blue sphere in the figure) enclosing a Weyl node is a quantized integer, defining the chiral charge of the Weyl fermion.}
\label{box1}
\end{figure*}

\end{document}